\documentclass[11pt,english,a4]{article}
\usepackage[latin1]{inputenc}
\usepackage{geometry}
\usepackage{hyperref}
\usepackage{cite}
\usepackage{babel}

\usepackage{amsmath}
\usepackage{amssymb}

\numberwithin{equation}{section}

\newcommand{\mc}[1]{\mathcal{#1}}
\newcommand{\eps}{\varepsilon}

\newcommand{\diag}{{\rm diag}\,}

\newcommand{\de} {{\rm d\hspace{-1.5pt}}}

\newcommand{\os}{\overset}
\newcommand{\us}{\underset}
\newcommand{\rem}[1]{}

\renewcommand{\le}[1]{\lefteqn{#1}}

\date{}
\title{On the Covariant Quantization of Type II Superstrings}

\author{Sebastian Guttenberg%
\thanks{basti@hep.itp.tuwien.ac.at
}, Johanna Knapp%
\thanks{knapp@hep.itp.tuwien.ac.at
}, Maximilian Kreuzer%
\thanks{kreuzer@hep.itp.tuwien.ac.at
} \vspace{0.4cm}\\
Institut für Theoretische Physik, Technische Universität Wien,\\Wiedner
Hauptstra\ss e 8-10, A-1040 Vienna, Austria}

\begin{document}
\enlargethispage*{2cm}
\newcommand{\norm}[1]{{\parallel #1 \parallel }}

\newcommand{\Partiell}[2]{\left( \frac{\partial #1 }{\partial #2 }\right) }

\newcommand{\funktional}[2]{\frac{\delta #1 }{\delta #2 }}

\newcommand{\total}[2]{\frac{\de #1 }{\de #2 }}

\newcommand{\Frac}[2]{\left( \frac{#1 }{#2 }\right) }

\newcommand{\partiell}[2]{\frac{\partial #1 }{\partial #2 }}

\newcommand{\Rh}{{R_{h}}}

\newcommand{\To}{\rightarrow }
 
\newcommand{\ket}[1]{|#1 >}

\newcommand{\bra}[1]{<#1 |}

\newcommand{\Ket}[1]{\left| #1 \right\rangle }

\newcommand{\Bra}[1]{\left\langle #1 \right| }
 
\newcommand{\braket}[2]{<#1 |#2 >}

\newcommand{\Braket}[2]{\Bra{#1 }\left. #2 \right\rangle }

\newcommand{\kom}[2]{[#1 ,#2 ]}

\newcommand{\Kom}[2]{\left[ #1 ,#2 \right] }

\newcommand{\abs}[1]{\mid #1 \mid }

\newcommand{\Abs}[1]{\left| #1 \right| }

\newcommand{\erw}[1]{\langle #1 \rangle }

\newcommand{\Erw}[1]{\left\langle #1 \right\rangle }

\newcommand{\bei}[2]{\left. #1 \right| _{#2 }}

\newcommand{\dann}{\Rightarrow }

\newcommand{\q}[1]{\underline{#1 }}

\newcommand{\hoch}[1]{{}^{#1 }}

\newcommand{\tief}[1]{{}_{#1 }}

\newcommand{\UB}[2]{\protect\underbrace{#1 }_{\le {#2 }}}

\newcommand{\OB}[2]{\protect\overbrace{#1 }^{\le {#2 }}}

\newcommand{\rightj}{j^{\! R}{}}

\newcommand{\rightJ}{J^{\! R}{}}

\newcommand{\rightTheta}{\Theta ^{\! R}{}}
 
\newcommand{\rightomega}{\omega ^{\! R}{}}

\newcommand{\rightK}{K^{\! R}{}}

\newcommand{\rightLambda}{\Lambda ^{\! R}{}}

\newcommand{\feld}{\varphi }

\newcommand{\antifeld}{\pi }

\newcommand{\fussend}{\diamond }

\newcommand{\mf}[1]{#1 }
\renewcommand{\mf}[1]{\mathfrak{#1}}
\newcommand{\bmcI}{}
\newcommand{\emcI}{}
 \newcommand{\bmc}{}
\newcommand{\emc}{}

\maketitle	

\begin{picture}(0,0)\unitlength=1mm\put(145,70){TUW--04--11}\end{picture}
\begin{abstract}
In a series of papers Grassi, Policastro, Porrati and van Nieuwenhuizen
have introduced a new method to covariantly quantize the GS-superstring
by constructing a resolution of the pure spinor constraint of Berkovits'
approach. Their latest version is based on a gauged WZNW model and
a definition of physical states in terms of relative cohomology groups.
We first put the off-shell formulation of the type II version of their
ideas into a chirally split form and directly construct the free action
of the gauged WZNW model, thus circumventing some complications of
the super group manifold approach to type II. Then we discuss the BRST 
charges that define the relative cohomology and the N=2 superconformal 
algebra. A surprising result is that nilpotency of the BRST charge 
requires the introduction of another quartet of ghosts.\end{abstract}
\tableofcontents{}
\newpage

\section{Introduction}

The classical Green Schwarz superstring%
\footnote{\( x^{m} \), \( \theta ^{\alpha } \), \( \hat{\theta }^{\hat{\alpha }} \)
are the coordinates of the target-superspace, where \( \theta ^{\alpha } \)
and \( \hat{\theta }^{\hat{\alpha }} \) are Majorana-Weyl, i.e. they
each have 16 real components.\( \quad \fussend  \)
}\begin{eqnarray}
\mc {L}^{GS} & = & -\frac{1}{2}\Pi _{\mu }^{m}\Pi ^{\mu }_{m}+\mc {L}_{WZ}\\
\Pi ^{m}_{\mu } & = & \partial _{\mu }x^{m}-i\theta \gamma ^{m}\partial _{\mu }\theta -i\hat{\theta }\gamma ^{m}\partial _{\mu }\hat{\theta }\\
\mc {L}_{WZ} & = & -i\eps ^{\mu \nu }\Pi _{\mu }^{m}\left( (\theta \gamma _{m}\partial _{\nu }\theta )-(\hat{\theta }\gamma _{m}\partial _{\nu }\hat{\theta })\right) -\eps ^{\mu \nu }(\theta \gamma _{m}\partial _{\mu }\theta )(\hat{\theta }\gamma ^{m}\partial _{\nu }\hat{\theta })
\end{eqnarray}
is covariant and manifestly spacetime supersymmetric. In this last
feature it differs from the RNS string, where space time supersymmetry
only comes in after the GSO projection. The problem for the Green Schwarz
string on the other hand is that a covariant quantization with the
standard BRST procedure does not work. The reason for this misery
is a set of 16 mixed first and second class constraints \( d_{z\alpha } \)
that cannot be split easily into first and second class type in a
covariant manner. The conjugate momenta \( p_{z\alpha } \)
of \( \theta ^{\alpha } \) can be entirely expressed by other phase
space variables. \( p_{z\alpha } \) minus these expressions
are thus constraints in the Hamiltonian formalism, namely \( d_{z\alpha } \).
Half of the constraints are first class and correspond to a fermionic
gauge symmetry, known as \( \kappa  \)-symmetry. Siegel \cite{Siegel:1986xj}
had the idea to make \( d_{z\alpha } \) part of a closed algebra
by adding the generators that arise via the Poisson bracket of \( d \)
with itself. At the quantum level this translates into the current
algebra
\begin{eqnarray}
id_{z\alpha }(z)id_{z\beta }(w) & \sim  & -2i\frac{\gamma ^{m}_{\alpha \beta }\Pi _{zm}(w)}{z-w}\\
id_{z\alpha }(z)\Pi _{zm}(w) & \sim  & -2\frac{\gamma _{m\, \alpha \beta }\partial \theta ^{\beta }(w)}{z-w}\\
\Pi _{zm}(z)\Pi _{zn}(w) & \sim  & -\frac{\eta _{mn}}{(z-w)^{2}}\\
id_{z\alpha }(z)\partial \theta ^{\beta }(w) & \sim  & -\frac{i\delta _{\alpha }^{\beta }}{(z-w)^{2}}
\end{eqnarray}
which corresponds to a chiral symmetry algebra of a free
action \begin{eqnarray}
\mc {L} & = & -\frac{1}{2}\partial x^{m}\bar{\partial }x_{m}+p_{z\alpha }\bar{\partial }\theta ^{\alpha }+\hat{p}_{\bar{z}\hat{\alpha }}\partial \hat{\theta }^{\hat{\alpha }}=\\
 & = & -\frac{1}{2}\Pi _{z}^{m}\Pi _{\bar{z}m}+\mc {L}_{WZ}+d_{z\alpha }\bar{\partial }\theta ^{\alpha }+\hat{d}_{\bar{z}\hat{\alpha }}\partial \hat{\theta }^{\hat{\alpha }}\\
d_{\mu \alpha } & \equiv  & p_{\mu \alpha }-(\gamma _{m}\theta )_{\alpha }\Big (i\partial _{\mu }x^{m}+\frac{1}{2}\theta \gamma ^{m}\partial _{\mu }\theta +\frac{1}{2}\hat{\theta }\gamma ^{m}\partial _{\mu }\hat{\theta }\Big )\\
\hat{d}_{\mu \hat{\alpha }} & \equiv  & \hat{p}_{\mu \hat{\alpha }}-(\gamma _{m}\hat{\theta })_{\hat{\alpha }}\Big (i\partial _{\mu }x^{m}+\frac{1}{2}\theta \gamma ^{m}\partial _{\mu }\theta +\frac{1}{2}\hat{\theta }\gamma ^{m}\partial _{\mu }\hat{\theta }\Big )
\end{eqnarray}
It has the interesting feature that it coincides with the Green-Schwarz
string, if one imposes the additional constraints \( d_{z\alpha }=0 \)
and \( \hat{d}_{\bar{z}\hat{\alpha }}=0 \) . In the free theory, \( d_{z\alpha } \)
is a priori not a Hamiltonian constraint, but the generator of a chiral
(not local) symmetry. The second class property of \( d_{z\alpha } \)
in the Green-Schwarz string is reflected in the free theory by the
fact that the OPE of \( d_{z\alpha } \) with itself does not form
a closed subalgebra. Berkovits \cite{Berkovits:2000fe} had the idea
to implement the constraint cohomologically with a BRST operator disregarding
its non-closure\begin{equation}
\label{BerkovitsBRST}
Q=-\oint \de zi\lambda ^{\alpha }d_{z\alpha }-\oint \de \bar{z}i\hat{\lambda }^{\hat{\alpha }}\hat{d}_{\bar{z}\hat{\alpha }}
\end{equation}
 where \( \lambda ^{\alpha } \) is a commuting ghost. For first class
constraints the BRST cohomology can be built, because the BRST operator
is nilpotent due to the closure of the algebra. For second class constraints,
however, the non-closure implies a lack of nilpotency of the BRST
operator. To overcome this problem, Berkovits put a constraint on
the ghost field \( \lambda  \) and \( \hat{\lambda } \), the so
called pure spinor constraint \begin{equation}
\lambda \gamma ^{m}\lambda =0
\end{equation}
 In a series of papers \cite{Nh:2001ug}\,--\!\!\cite{Nh:2004cz}
Grassi, Policastro, Porrati and van Nieuwenhuizen have removed this
extra constraint by adding additional ghost variables. They realized
in \cite{Nh:2003kq} that their theory has the stucture of a gauged
WZNW model with the complete diagonal subgroup gauged. It is based
on the chiral algebra above (note that we changed the sign of 
\( p_{z\alpha } \)
and \( d_{z\alpha } \) in our definitions as compared to \cite{Nh:2003kq}).
A current can be set to zero by gauging the corresponding symmetry
and thus making it a first class constraint. However, \( d_{z\alpha } \)
does not form a subalgebra and thus cannot be gauged on its own. So
if one starts gauging \( d_{z\alpha } \) and tries to make the resulting
BRST-operator (\ref{BerkovitsBRST}) nilpotent by adding further ghosts,
one automatically arrives at a BRST operator that corresponds to a
theory where also \( \Pi _{zm} \) and \( \partial \theta ^{\alpha } \)
are gauged (see e.g. \cite[p.7]{Nh:2003cm} or
\cite[p.4]{Nh:2003kq}). But this also removes the physical fields from cohomology.
Therefore a grading was introduced by hand to get
the correct cohomology. In a recent beautiful paper \cite{Nh:2004cz} this 
procedure was replaced by a new approach to the conversion of second class
constraints into first class, which was worked out
for the gauging of a coset of a (simple) Lie algebra.
The idea, which may work for a more general set of constraints
that generate a first class algebra, is to
gauge the complete algebra and later undo
the gauging of the unwanted constraints by building a relative cohomology
with respect to a second BRST operator. In the present paper we discuss
this program for the type II string with some modifications.

Despite of the elegance of the group supermanifold approach, there are some 
puzzling points about the WZNW action:

\begin{itemize}
\item For the heterotic string one starts with a chiral algebra and gets
from the WZNW model a chiral as well as an antichiral algebra. Somehow
one has to get rid of the antichiral one.
\item For the type II string one starts with a chiral and an antichiral 
algebra.
Both of them double and, more severely, because of the central extension of
the supersymmetry algebra, the Jacobi identity forces one 
to introduce an additional generator with two spinor indices.
Thus it has not been possible yet for us to produce a WZNW
action for the type II string.
\end{itemize}
Although we hope that those problems can be overcome, we shortcut the
procedure by directly constructing the free field version of the gauged
WZNW action. \\
The paper is organized as follows:

\begin{itemize}
\item In section \ref{gauging} we start with the free field action,
discuss its off-shell symmetry algebra generated by the 
current \( d_{z\alpha } \)
and gauge it to turn \( d_{z\alpha } \) into a constraint. Before
actually gauging the algebra via the Noether procedure, we 
make it close off-shell. To this end we introduce auxiliary fields
\( P_{zm} \) and \( P_{\bar{z}m} \). There still remain double poles
in the current algebra, which cause trouble in the gauging procedure.
They can be eliminated by doubling all fields,
as it was done in \cite{Nh:2003kq},
in order to establish nilpotent BRST transformations. Gauge fixing
leads to the BRST-transformations as they are given in \cite{Nh:2003kq}.
\item In \ref{secondBRST} we follow the ideas in \cite{Nh:2004cz} to 
define a relative cohomology, whose purpose will be to undo the gauging
of part of the algebra. A nilpotent second BRST operator can be
found for the case of arbitrary generators of a first class algebra,
but the construction presumably has to be modified for the superstring.

\item In \ref{operatoralgebra} we review the operator algebra presented in 
\cite{Nh:2003kq}. It is
straightforwardly extended by the new fields that were introduced in
the context of the second BRST operator.
\item In \ref{diffeomorphismus}, we have a closer look at the
final BRST operator proposed in \cite{Nh:2003kq}, which includes
diffeomorphism invariance by adding a topological ghost quartet. We
come to the conclusion that this operator has to be modified via a
second quartet of ghost fields in order to become nilpotent.
However, we have not found a suitable form of the second BRST
      operator at this final stage, because its nilpotency is spoiled by
      additional terms that are required by diffeomorphism invariance.
      In this situation one can either do without diffeomorphisms 
      (but keep the $N=2$ algebra), or return to the grading 
      prescription or a modification thereof.
\end{itemize}
We also want to mention some other recent work that is based on
the Berkovits approach \cite{Che,AK,Matone:2002ft}.

\section{Gauging the Current Algebra}

\label{gauging}We start with the free field Lagrangian%
\footnote{For a detailed description of our conventions, including the 
definition
of the chiral projector \( P^{\mu \nu } \), we refer to the 
appendix.\( \quad \fussend  \)
}\begin{eqnarray}
\mc {L} & \equiv  & -\frac{1}{2}\partial x^{m}\bar{\partial 
}x_{m}+p_{z\alpha }\bar{\partial }\theta ^{\alpha }
+\hat{p}_{\bar{z}\hat{\alpha }}\partial \hat{\theta }^{\hat{\alpha }}
=\label{freefield} \\
 & = & -\frac{1}{2}\Pi _{z}^{m}\Pi _{\bar{z}m}+\mc {L}_{WZ}+d_{z\alpha }
\bar{\partial }\theta ^{\alpha }+\hat{d}_{\bar{z}\hat{\alpha }}\partial 
\hat{\theta }^{\hat{\alpha }}=\label{nearGS} \\
 & = & -\frac{1}{2}\Pi _{\mu }^{m}\Pi ^{\mu }_{m}+\mc {L}_{WZ}+P^{\mu \nu }
d_{\mu \alpha }\partial _{\nu }\theta ^{\alpha }+\bar{P}^{\mu \nu }
\hat{d}_{\mu \hat{\alpha }}\partial _{\nu }\hat{\theta }^{\hat{\alpha }}\\
\Pi ^{m}_{\mu } & \equiv  & \partial _{\mu }x^{m}-i\theta \gamma ^{m}
\partial _{\mu }\theta -i\hat{\theta }\gamma ^{m}\partial _{\mu }
\hat{\theta }\\
d_{\mu \alpha } & \equiv  & p_{\mu \alpha }-(\gamma _{m}\theta )_{\alpha }
\Big (i\partial _{\mu }x^{m}+\frac{1}{2}\theta \gamma ^{m}\partial _{\mu }
\theta +\frac{1}{2}\hat{\theta }\gamma ^{m}\partial _{\mu }\hat{\theta }
\Big )\\
\hat{d}_{\mu \hat{\alpha }} & \equiv  & \hat{p}_{\mu \hat{\alpha }}
-(\gamma _{m}\hat{\theta })_{\hat{\alpha }}\Big (i\partial _{\mu }x^{m}
+\frac{1}{2}\theta \gamma ^{m}\partial _{\mu }\theta +\frac{1}{2}
\hat{\theta }\gamma ^{m}\partial _{\mu }\hat{\theta }\Big )\\
\mc {L}_{WZ} & \equiv  & -i\eps ^{\mu \nu }\Pi _{\mu }^{m}\left( (\theta 
\gamma _{m}\partial _{\nu }\theta )-(\hat{\theta }\gamma _{m}\partial 
_{\nu }\hat{\theta })\right) -\eps ^{\mu \nu }(\theta \gamma _{m}\partial 
_{\mu }\theta )(\hat{\theta }\gamma ^{m}\partial _{\nu }\hat{\theta })
\end{eqnarray}
 From the first line (\ref{freefield}) to the second line (\ref{nearGS})
we reexpressed the Lagrangian in terms of supersymmetric objects%
\footnote{Supersymmetry acts on the superspace coordinates \( x^{m} \), 
\( \theta ^{\alpha } \)
and \( \hat{\theta }^{\hat{\alpha }} \) as follows \begin{eqnarray*}
\delta _{\eps }\theta ^{\alpha } & = & \eps ^{\alpha }\\
\delta _{\eps }\hat{\theta }^{\hat{\alpha }} & 
= & \hat{\eps }^{\hat{\alpha }}\\
\delta _{\eps }x^{m} & 
= & i(\eps \gamma ^{m}\theta )+i(\hat{\eps }\gamma ^{m}\hat{\theta })
\end{eqnarray*}
\[
\dann \quad \delta _{\eps }\partial \theta ^{\alpha }=\delta _{\eps }
\partial \hat{\theta }^{\hat{\alpha }}=\delta _{\eps }\Pi ^{m}_{\mu }=0\]
 The transformations of \( p_{\mu \alpha } \) and \( \hat{p}_{\mu 
\hat{\alpha }} \)
under supersymmetry \( \delta _{\eps } \) are defined in such a way
that \( d_{\mu \alpha } \) is SUSY inert.\begin{eqnarray*}
\delta _{\eps }p_{\mu \alpha } & = & (\gamma _{m}\eps )_{\alpha }\Big 
(i\partial _{\mu }x^{m}+\frac{1}{2}\theta \gamma ^{m}\partial _{\mu }\theta 
+\frac{1}{2}\hat{\theta }\gamma ^{m}\partial _{\mu }\hat{\theta }\Big )
-\frac{1}{2}(\gamma _{m}\theta )_{\alpha }\Big (\eps \gamma ^{m}\partial 
_{\mu }\theta +\hat{\eps }\gamma ^{m}\partial _{\mu }\hat{\theta }\Big )\\
\delta _{\eps }\hat{p}_{\mu \hat{\alpha }} & = & (\gamma _{m}\hat{\eps })_{\hat{\alpha }}\Big (i\partial _{\mu }x^{m}+\frac{1}{2}\theta \gamma ^{m}\partial _{\mu }\theta +\frac{1}{2}\hat{\theta }\gamma ^{m}\partial _{\mu }\hat{\theta }\Big )-\frac{1}{2}(\gamma _{m}\hat{\theta })_{\hat{\alpha }}\Big (\eps \gamma ^{m}\partial _{\mu }\theta +\hat{\eps }\gamma ^{m}\partial _{\mu }\hat{\theta }\Big )
\end{eqnarray*}
\[
\dann \quad \delta _{\eps }d_{\mu \alpha }=\delta _{\eps }\hat{d}_{\mu \hat{\alpha }}=0\]
\( \mc {L}_{WZ} \) is not invariant itself, but transforms into a
total divergence.\[
\delta _{\eps }\int \mc {L}_{WZ}=0\]
We want to have a theory in the end which is manifestly supersymmetric.
Therefore the BRST differential to be constructed should anticommute
with supersymmetry. This will be a leading thought in the \(\rm following.\,\,\fussend  \)
}. In (\ref{nearGS}) it becomes obvious that the free field action
turns into the Green Schwarz action for the type II superstring if
\( d_{z\alpha }=0 \) and \( \hat{d}_{\bar{z}\hat{\alpha }}=0 \)
are put as extra constraints. 
The free field equations (\( \partial \bar{\partial }x^{m}=\bar{\partial }\theta ^{\alpha }=\bar{\partial }p_{z\alpha }=\partial 
\hat{\theta }^{\hat{\alpha }}=\partial \hat{p}_{\bar{z}\hat{\alpha }}=0 \))
imply that \( d_{z\alpha } \) is a conserved 
chiral and \( \hat{d}_{\bar{z}\hat{\alpha }} \)
is a conserved antichiral Noether current\begin{eqnarray}
\bar{\partial }d_{z\alpha } & \stackrel{\textrm{eom}}{=} & 0,\qquad \partial \hat{d}_{\bar{z}\hat{\alpha }}\stackrel{\textrm{eom}}{=}0
\end{eqnarray}
They thus belong to symmetries of the free field action. As discussed
in the introduction, we want to turn those currents into constraints
by gauging the symmetries. Instead of the imaginary objects \( d_{z\alpha } \)
and \( \hat{d}_{\bar{z}\alpha } \), we take \( id_{z\alpha } \)
and \( i\hat{d}_{\bar{z}\hat{\alpha }} \) as currents. Although we
want to discuss the off-shell symmetries, the symmetry algebra (on-shell)
can be most neatly written down in terms of operator products. We
will do so, in order to get an idea, how we have to proceed. On the
operator level, \( id_{z\alpha } \) is part of the following operator
algebra: \begin{eqnarray}
id_{z\alpha }(z)id_{z\beta }(w) & \sim  & -2i\frac{\gamma ^{m}_{\alpha \beta }\Pi _{zm}(w)}{z-w}\\
id_{z\alpha }(z)\Pi _{zm}(w) & \sim  & -2\frac{\gamma _{m\, \alpha \beta }\partial \theta ^{\beta }(w)}{z-w}\\
\Pi _{zm}(z)\Pi _{zn}(w) & \sim  & -\frac{\eta _{mn}}{(z-w)^{2}}\\
id_{z\alpha }(z)\partial \theta ^{\beta }(w) & \sim  & -\frac{i\delta _{\alpha }^{\beta }}{(z-w)^{2}}
\end{eqnarray}
Here it becomes obvious that \( d \) does not form a closed subalgebra.
This reflects the fact that it corresponds to second class constraints
in the Green Schwarz string. Instead, the remaining currents \( \Pi _{zm} \)
and \( \partial \theta ^{\alpha } \) form a (centrally extended) subalgebra. 
Following
the instructions of \cite{Nh:2004cz}, we thus have to gauge the complete
algebra and later build the relative cohomology of the BRST operator
with respect to a second one which will undo the gauging of \( \Pi  \)
and \( \partial \theta  \).\\
Let us rewrite the algebra in a more general notation:%
\footnote{Here we have a different sign in the definition of the structure constants
as compared to \cite{Nh:2003kq}. The reason is that in our conventions
the algebra \[
J_{M_{1}}(z)J_{M_{2}}(w)\sim -\frac{J_{M_{3}}f^{M_{3}}\tief{M_{1}M_{2}}}{z-w}-\frac{\mc {H}_{M_{1}M_{2}}}{(z-w)^{2}}\]
corresponds to a Poisson bracket algebra\begin{eqnarray}
\left\{ J_{M_{1}}(\sigma ^{-}),J_{M_{2}}(\sigma '^{-})\right\}  & = & 4\pi J_{M_{3}}f^{M_{3}}\tief{M_{1}M_{2}}\delta (\sigma ^{1}-\sigma '^{1})-8\pi \mc {H}_{M_{1}M_{2}}\partial _{1}\delta (\sigma ^{1}-\sigma '^{1})\nonumber 
\end{eqnarray}
This is excatly the classical algebra that one gets from a WZNW-model
with level \( n=-2 \) and with currents\begin{eqnarray}
J_{M} & \equiv  & \Erw{T_{M}\, ,\, g^{-1}\partial g}\quad \iff \quad g^{-1}\partial g=T_{M}J^{M}\nonumber \\
\textrm{with }\left[ T_{M_{1}},T_{M_{2}}\right]  & = & T_{M_{3}}f^{M_{3}}\tief{M_{1}M_{2}}\nonumber 
\end{eqnarray}
Therefore we think that the sign of the structure constants in \cite{Nh:2003kq},
eq. (2.3) is not consistent with the choice \( \left[ T_{M_{1}},T_{M_{2}}\right] =T_{M_{3}}f^{M_{3}}\tief{M_{1}M_{2}} \).
That explains the need for a non-invariant metric \( H_{MN} \) in
\cite{Nh:2003kq} to pull the indices of the currents 
once.\( \quad \fussend  \)
} \begin{eqnarray}
J_{M_{1}}(z)J_{M_{2}}(w) & \sim  & -\frac{J_{M_{3}}f^{M_{3}}\tief{M_{1}M_{2}}}{z-w}-\frac{\mc {H}_{M_{1}M_{2}}}{(z-w)^{2}}\label{algebra} \\
J_{zM}\equiv J_{M} & \equiv  & (J_{m},J_{\alpha },J_{\q{\alpha }})\equiv (\Pi _{zm},id_{z\alpha },\partial \theta ^{\alpha })\\
\mc {H}_{M_{1}M_{2}} & \equiv  & \left( \begin{array}{ccc}
\eta _{m_{1}m_{2}} & 0 & 0\\
0 & 0 & \mc {H}_{\alpha _{1}\q{\alpha }_{2}}=i\delta _{\alpha _{1}}^{\alpha _{2}}\\
0 & \mc {H}_{\q{\alpha }_{1}\alpha _{2}}=-i\delta _{\alpha _{2}}^{\alpha _{1}} & 0
\end{array}\right) \\
f^{m_{3}}\tief{\alpha _{1}\alpha _{2}} & \equiv  & 2i\gamma ^{m}_{\alpha \beta }=f^{m_{3}}\tief{\alpha _{2}\alpha _{1}}\\
f^{\q{\alpha }_{3}}\tief{\alpha _{1}m_{2}} & \equiv  & 2\gamma _{m_{2}\, \alpha _{1}\alpha _{3}}=-f^{\q{\alpha }_{3}}\tief{m_{2}\alpha _{1}}
\end{eqnarray}
We can use the metric and its graded inverse\begin{eqnarray}
\mc {H}^{M_{1}M_{2}} & \equiv  & \left( \begin{array}{ccc}
\eta ^{m_{1}m_{2}} & 0 & 0\\
0 & 0 & \mc {H}^{\alpha _{1}\q{\alpha }_{2}}=-i\delta ^{\alpha _{1}}_{\alpha _{2}}\\
0 & \mc {H}^{\q{\alpha }_{1}\alpha _{2}}=i\delta ^{\alpha _{2}}_{\alpha _{1}} & 0
\end{array}\right) 
\end{eqnarray}
to pull indices:\begin{equation}
J^{M}\equiv (J^{m},J^{\alpha },J^{\q{\alpha }})=J_{N}\mc {H}^{NM}=(\Pi ^{m}_{z},i\partial \theta ^{\alpha },d_{z\alpha })
\end{equation}
For the antichiral currents \( \hat{J}_{\bar{z}M}\equiv \hat{J}_{\hat{M}}\equiv (\hat{J}_{m},\hat{J}_{\hat{\alpha }},\hat{J}_{\q{\hat{\alpha }}})\equiv (\Pi _{\bar{z}m},i\hat{d}_{\bar{z}\alpha },\bar{\partial }\hat{\theta }^{\hat{\alpha }}) \),
we define a metric \( \mc {\hat{H}}_{\hat{M}\hat{N}} \) and structure
constants \( \hat{f}^{\hat{K}}\tief{\hat{M}\hat{N}} \) that are numerically
equal to those above.

\subsection{The Form of the Transformations to be Gauged}

Let us also define a generalized notation for the field content
we have so far%
\footnote{We call those fields {}``matter fields'' in the following, in order
to distinguish them from ghost fields and auxiliary fields, which
we are going to introduce later.\( \quad \fussend  \)
}\begin{equation}
\phi ^{\mc {A}}\equiv (\phi ^{m},\phi ^{\alpha },\phi ^{\q{\alpha }},\phi ^{\hat{\alpha }},\phi ^{\q{\hat{\alpha }}})\equiv (x^{m},\theta ^{\alpha },p_{z\alpha },\hat{\theta }^{\hat{\alpha }},\hat{p}_{\bar{z}\hat{\alpha }})
\end{equation}
Introduce transformation parameters \( \omega ^{M} \) and \( \hat\omega ^{\hat{M}} \)
with Lie-Algebra indices\begin{eqnarray}
\omega ^{M}(z) & \equiv  & (\omega ^{m},\omega ^{\alpha },\omega ^{\q{\alpha }})=(\omega ^{m},\omega ^{\alpha },-i\omega _{\alpha })\\
\omega _{M}(z) & \equiv  & (\omega _{m},\omega _{\alpha },\omega _{\q{\alpha }})=(\omega _{m},\omega _{\alpha },-i\omega ^{\alpha })\\
\hat{\omega }^{\hat M}(\bar{z}) & \equiv  & (\hat{\omega }^{m},\hat{\omega }^{\hat{\alpha }},\hat{\omega }^{\q{\hat{\alpha }}})=(\hat{\omega }^{m},\hat{\omega }^{\hat{\alpha }},-i\hat{\omega }_{\hat{\alpha }})\\
\hat{\omega }^{\hat M}(\bar{z}) & \equiv  & (\hat{\omega }_{m},\hat{\omega }_{\hat{\alpha }},\hat{\omega }_{\q{\hat{\alpha }}})=(\hat{\omega }_{m},\hat{\omega }_{\hat{\alpha }},-i\hat{\omega }^{\hat{\alpha }})
\end{eqnarray}
Noether's theorem tells us that one gets the current by a local variation
of the fields\begin{eqnarray}
\delta _{\omega }S & = & \int \funktional{S}{\phi ^{\mc {A}}}\delta _{\omega }\phi ^{\mc {A}}\stackrel{!}{=}\int J_{M}\bar{\partial }\omega ^{M}=-\int \bar{\partial }J_{M}\omega ^{M}\label{Noether}
\end{eqnarray}
with\begin{equation}
\funktional{S}{\phi ^{\mc {A}}}=(\partial \bar{\partial }x^{m},-\bar{\partial }p_{z\alpha },-\bar{\partial }\theta ^{\alpha },-\partial \hat{p}_{\bar{z}\hat{\alpha }},-\partial \hat{\theta }^{\hat{\alpha }})
\end{equation}
Given the currents, we thus can read off the transformations according
to (\ref{Noether})\footnote{In this and several future calculations, one has to use the Fierz identity \( \gamma ^{\, }_{m\, \alpha (\beta }\gamma ^{m}_{\gamma \delta )}=0 \)
.\( \quad \fussend  \)
}\begin{eqnarray}
\le {-\int \bar{\partial }J_{M}\omega ^{M}\; =\; \int -\bar{\partial }\Pi _{zm}\omega ^{m}-\bar{\partial }id_{z\alpha }\omega ^{\alpha }+i\bar{\partial }\partial \theta ^{\alpha }\omega _{\alpha }=} &  & \\
 & = & \int \bar{\partial }\partial x_{m}\Big \{-\omega ^{m}+(\omega \gamma ^{m}\theta )\Big \}-\bar{\partial }p_{z\alpha }\left( i\omega ^{\alpha }\right) +\nonumber \\
 &  & -\bar{\partial }\theta ^{\alpha }\Big \{i\partial \omega _{\alpha }-2i\omega ^{m}(\gamma _{m}\partial \theta )_{\alpha }-i\partial \omega ^{m}(\gamma _{m}\theta )_{\alpha }+(\gamma _{m}\omega )_{\alpha }\partial x^{m}+\nonumber \\
 &  & \qquad \quad +\frac{3i}{2}(\omega \gamma _{m}\theta )(\gamma ^{m}\partial \theta )_{\alpha }+\frac{i}{2}(\partial \omega \gamma _{m}\theta )(\gamma ^{m}\theta )_{\alpha }\Big \}+\nonumber \\
 &  & -\partial \hat{\theta }^{\hat{\alpha }}\Big \{-i(\gamma _{m}\hat{\theta })_{\alpha }\bar{\partial }\omega ^{m}+\frac{i}{2}(\bar{\partial }\omega \gamma _{m}\theta )(\gamma ^{m}\hat{\theta })_{\hat{\alpha }}\Big \}
\end{eqnarray}
 There is, however, an arbitrariness: Adding \( (\gamma _{m}\mu )_{\alpha }(\hat{\theta }\gamma ^{m}\partial \hat{\theta }) \)
with some parameter \( \mu ^{\alpha } \) to the bracket behind \( \bar{\partial }\theta  \)
and adding at the same time a term \( (\mu \gamma _{m}\bar{\partial }\theta )(\gamma ^{m}\hat{\theta })_{\hat{\alpha }} \)
to the bracket behind \( \partial \hat{\theta }^{\hat{\alpha }} \)
does not change anything. This is a special form of a \emph{trivial
gauge transformation}%
\footnote{A trivial gauge transformation (see e.g. \cite[p.69]{Henneaux:1992ig})
is of the form \[
\delta \phi ^{M}=(-)^{N}A^{MN}\funktional{S}{\phi ^{N}}\]
 with \( A^{MN} \) graded antisymmetric. It is a local symmetry that
is present for any theory with more than one field, but it does not
imply a gauge freedom: \[
\delta S=\funktional{S}{\phi ^{M}}\delta \phi ^{M}=(-)^{N}\funktional{S}{\phi ^{M}}A^{MN}\funktional{S}{\phi ^{N}}=0\]
 In the present case we have \begin{eqnarray*}
A^{\q{\alpha }\q{\hat{\beta }}} & = & (\gamma _{m}\hat{\mu })_{\hat{\beta }}(\gamma ^{m}\theta )_{\alpha }+(\gamma _{m}\mu )_{\alpha }(\gamma ^{m}\hat{\theta })_{\hat{\beta }}\\
A^{\q{\hat{\alpha }}\q{\beta }} & = & (\gamma _{m}\hat{\mu })_{\hat{\alpha }}(\gamma ^{m}\theta )_{\beta }+(\gamma _{m}\mu )_{\beta }(\gamma ^{m}\hat{\theta })_{\hat{\alpha }}\quad \fussend 
\end{eqnarray*}

} of \( p_{z\alpha } \) and \( \hat{p}_{\bar{z}\hat{\alpha }} \).
Trivial transformations do not change the currents. We have started
with supersymmetric currents and would have expected that they correspond
to transformations which commute with supersymmetry. But this is only
true up to trivial gauge transformations. We do want transformations
that commute with supersymmetry because those transformations will
later (after gauging and gauge fixing) turn into BRST transformations.
If we want an off-shell formalism which is manifestly supersymmetric,
BRST should commute with supersymmetry. We therefore choose the transformations
in such a way that this is fulfilled. We further add right now the
transformations which correspond to the antiholomorphic currents:%
\footnote{We also could have derived all this using operator product expansions
(or classically via the Poisson-bracket). In the quantized theory,
the currents are generators of the symmetry transformations via the
operator product:\begin{eqnarray}
\delta _{\omega }\phi ^{N}(w) & = & \us {z\To w}{\rm Res}\: J_{M}(z)\omega ^{M}(z)\phi ^{N}(w)\nonumber 
\end{eqnarray}
The basic OPEs of the free field theory in our conventions are \begin{eqnarray*}
\partial x^{m}(z)\partial x^{n}(w) & \sim  & -\frac{\eta ^{mn}}{(z-w)^{2}}\\
p_{z\alpha }(z)\theta ^{\beta }(w) & \sim  & \frac{-\delta ^{\beta }_{\alpha }}{z-w}\; \sim \; d_{z\alpha }(z)\theta ^{\beta }(w)
\end{eqnarray*}
However, the resulting transformations would not contain a priori
the terms \( -i(\gamma _{m}\hat{\theta })_{\alpha }\bar{\partial }\omega ^{m}+\frac{i}{2}(\bar{\partial }\omega \gamma _{m}\theta )(\gamma ^{m}\hat{\theta })_{\hat{\alpha }} \)
in \( \delta \hat{p}_{\bar{z}\hat{\alpha }} \) which are necessary
to produce the right off-shell currents. In addition, the same considerations
about supersymmetry and trivial transformations have to be done there,
too.\( \quad \fussend  \)
}\begin{eqnarray}
\delta x^{m} & = & -\omega ^{m}+(\omega \gamma ^{m}\theta )-\hat{\omega }^{m}+(\hat{\omega }\gamma ^{m}\hat{\theta })\label{xtrafo} \\
\delta \theta ^{\alpha } & = & i\omega ^{\alpha }\label{thetatrafo} \\
\delta p_{z\alpha } & = & i\partial \omega _{\alpha }-2i\omega ^{m}(\gamma _{m}\partial \theta )_{\alpha }-i\partial \omega ^{m}(\gamma _{m}\theta )_{\alpha }+\nonumber \\
 &  & +(\gamma _{m}\omega )_{\alpha }\partial x^{m}+\frac{3i}{2}(\omega \gamma _{m}\theta )(\gamma ^{m}\partial \theta )_{\alpha }+\frac{i}{2}(\partial \omega \gamma _{m}\theta )(\gamma ^{m}\theta )_{\alpha }+ \\ \nonumber
 &  & -\frac{3i}{2}(\gamma _{m}\omega )_{\alpha }(\hat{\theta }\gamma ^{m}\partial \hat{\theta })-i\partial \hat{\omega }_{m}(\gamma ^{m}\theta )_{\alpha }+\frac{3i}{2}(\hat{\omega }\gamma _{m}\partial \hat{\theta })(\gamma ^{m}\theta )_{\alpha }+\frac{i}{2}(\partial \hat{\omega }\gamma ^{m}\hat{\theta })(\gamma _{m}\theta )_{\alpha }\\
\delta \hat{\theta }^{\hat{\alpha }} & = & i\hat{\omega }^{\hat{\alpha }}\label{thetahattrafo} \\
\delta \hat{p}_{\bar{z}\hat{\alpha }} & = & i\bar{\partial }\hat{\omega }_{\hat{\alpha }}-2i\hat{\omega }_{m}(\gamma ^{m}\bar{\partial }\hat{\theta })_{\hat{\alpha }}-i\bar{\partial }\hat{\omega }_{m}\, (\gamma ^{m}\hat{\theta })_{\hat{\alpha }}+\nonumber \\
 &  & +(\gamma _{m}\hat{\omega })_{\hat{\alpha }}\bar{\partial }x^{m}+\frac{3i}{2}(\hat{\omega }\gamma _{m}\hat{\theta })(\gamma ^{m}\bar{\partial }\hat{\theta })_{\hat{\alpha }}+\frac{i}{2}(\bar{\partial }\hat{\omega }\, \gamma _{m}\hat{\theta })\left( \gamma ^{m}\hat{\theta }\right) _{\hat{\alpha }}+\\ \nonumber
 &  & -\frac{3i}{2}(\gamma _{m}\hat{\omega })_{\hat{\alpha }}(\theta \gamma ^{m}\bar{\partial }\theta )-i\bar{\partial }\omega ^{m}(\gamma _{m}\hat{\theta })_{\alpha }+\frac{3i}{2}(\omega \gamma _{m}\bar{\partial }\theta )(\gamma ^{m}\hat{\theta })_{\hat{\alpha }}+\frac{i}{2}(\bar{\partial }\omega \gamma _{m}\theta )(\gamma ^{m}\hat{\theta })_{\hat{\alpha }}\label{phuttrafo} 
\end{eqnarray}
 It becomes obvious that these transformations commute with supersymmetry,
when one varies supersymmetric objects. The results should again be
supersymmetric, which is indeed the case:\begin{eqnarray}
\delta \Pi _{zm} & = & -\partial \omega _{m}+2(\omega \gamma _{m}\partial \theta )-\partial \hat{\omega }_{m}+2(\hat{\omega }\gamma _{m}\partial \hat{\theta })\label{Pitrafo} \\
\delta id_{z\alpha } & = & -\partial \omega _{\alpha }+2i(\gamma _{m}\omega )_{\alpha }\Pi _{z}^{m}+2\omega ^{m}(\gamma _{m}\partial \theta )_{\alpha }\label{dtrafo} \\
\delta \partial \theta ^{\alpha } & = & i\partial \omega ^{\alpha }\\
\delta \Pi _{\bar{z}m} & = & -\bar{\partial }\omega _{m}+2(\omega \gamma _{m}\bar{\partial }\theta )-\bar{\partial }\hat{\omega }_{m}+2(\hat{\omega }\gamma _{m}\bar{\partial }\hat{\theta })\\
\delta i\hat{d}_{\bar{z}\hat{\alpha }} & = & -\bar{\partial }\hat{\omega }_{\hat{\alpha }}+2i(\gamma _{m}\hat{\omega })_{\hat{\alpha }}\Pi _{\bar{z}}^{m}+2\hat{\omega }^{m}(\gamma _{m}\bar{\partial }\hat{\theta })_{\hat{\alpha }}\label{dhuttrafo} \\
\delta \bar{\partial }\hat{\theta }^{\hat{\alpha }} & = & i\bar{\partial }\hat{\omega }^{\hat{\alpha }}
\end{eqnarray}
In the condensed notation, this can be written as\begin{eqnarray}
\delta J_{M} & = & -\partial \omega _{M}+J_{P}f^{P}\tief{MN}\omega ^{N}+(\hat{\omega }-\textrm{terms})\\
\delta \hat{J}_{\hat{M}} & = & -\bar{\partial }\hat{\omega }_{\hat M}+\hat{J}_{\hat{P}}\hat{f}^{\hat{P}}\tief{\hat{M}\hat{N}}\hat{\omega }^{\hat{N}}+(\omega -\textrm{terms})
\end{eqnarray}
The fact that the transformation of \( J_{M} \) contains \( \hat{\omega } \)-terms
reflects the non-closure of the off-shell algebra of the transformations.

\subsection{Closing the Off-Shell-Algebra}

In order to arrive at an off-shell nilpotent BRST differential, one
needs an off-shell closed gauge algebra. The commutators of the transformations
generated by \( \Pi _{zm} \), \( id_{z\alpha } \), \( \partial \theta ^{\alpha } \),
\( \Pi _{\bar{z}m} \), \( i\hat{d}_{\bar{z}\hat{\alpha }} \) and
\( \bar{\partial }\hat{\theta }^{\hat{\alpha }} \) generate further
transformations which we have not yet included in the algebra. This
becomes evident only when a commutator of transformations is acting
on \( d \) or \( \hat{d} \). \begin{eqnarray}
\kom{\delta _{1}}{\delta _{2}}d_{z\alpha } & = & 2\partial \left( (\gamma _{m}\omega _{1})_{\alpha }\omega _{2}^{m}-\omega _{1}^{m}(\gamma _{m}\omega _{2})_{\alpha }\right) +4(\gamma _{m}\partial \theta )_{\alpha }(\omega _{1}\gamma ^{m}\omega _{2})+ \\ \nonumber
 &  & +2(\gamma _{m}\omega _{2})_{\alpha }\left( -\partial \hat{\omega }_{1}^{m}+2(\hat{\omega }_{1}\gamma ^{m}\partial \hat{\theta })\right) -2(\gamma _{m}\omega _{1})_{\alpha }\left( -\partial \hat{\omega }_{2}^{m}+2(\hat{\omega }_{2}\gamma ^{m}\partial \hat{\theta })\right) 
\end{eqnarray}
The first two terms correspond to transformations with parameter \( \omega _{\alpha } \)
and \( \omega ^{m} \) respectively. The last two terms show the non-closure
of the algebra. Furthermore, \( 2(\gamma _{m}\omega _{2})_{\alpha }\partial \hat{\omega }_{1}^{m} \)
and \( 2(\gamma _{m}\omega _{1})_{\alpha }\partial \hat{\omega }_{2}^{m} \)
correspond to pure shifts in \( d_{z\alpha } \) or equivalently in
\( p_{z\alpha } \). The holomorphic current for this transformation
is \( \theta ^{\alpha } \) which is not supersymmetric and thus would
spoil commutativity of BRST and SUSY. The reason for these terms to
show up, is the \( \Pi _{zm} \) in the transformation of \( d_{z\alpha } \)
(\ref{dtrafo}). The transformation of \( \Pi _{zm} \) (\ref{Pitrafo})
contains (in contrast to the heterotic case) also hatted variables. 

In \cite[p.11]{Nh:2002xf}, a nilpotent BRST transformation of \( d_{z\alpha } \)
was achieved by introducing an auxiliary variable \( P_{0}^{m} \)
and rewriting the transformations in terms of \( \partial _{1} \)-derivatives
only. We use a similar ansatz to get a closed off-shell algebra (this
is of course closely related to BRST-nilpotency). However, to keep the
manifest chiral split, we introduce two such variables. 

The symmetry algebra closes as long as we consider only the transformations
with parameter \( \omega ^{M} \) or only transformations with parameter
\( \hat{\omega }^{\hat{M}} \). This would correspond to the transformations
in a chiral or antichiral heterotic case. In order to get rid of the
terms in question, one can thus introduce auxiliary fields \( P_{z} \)
and \( P_{\bar{z}} \) in the transformation of \( d_{z} \) and \( \hat{d}_{\bar{z}\hat{\alpha }} \)
that transform as \( \Pi _{z} \) and \( \Pi _{\bar{z}} \) do in
the chiral or antichiral heterotic case respectively:\begin{eqnarray}
\tilde{\delta }d_{z\alpha } & = & i\partial \omega _{\alpha }+2(\gamma _{m}\omega )_{\alpha }P_{z}^{m}-2i\omega ^{m}(\gamma _{m}\partial \theta )_{\alpha }=\\
 & = & \delta d_{z\alpha }-2(\gamma _{m}\omega )_{\alpha }\left( \Pi _{z}^{m}-P_{z}^{m}\right) 
\end{eqnarray}
\begin{eqnarray}
\tilde{\delta }P_{z}^{m} & \stackrel{!}{=} & -\partial \omega ^{m}+2(\omega \gamma ^{m}\partial \theta )
\end{eqnarray}
 Equivalently for \( \hat{d}_{\bar{z}\hat{\alpha }} \) and \( P_{\bar{z}m} \).
These transformations are still supersymmetric within type II if \( P \)
is defined to be SUSY-inert. At the same time we have to guarantee
that the equations of motions of the matter fields do not change,
that on shell \( P \) coincides with \( \Pi  \) and that the invariance
of the action under the {}``global'' transformations (holomorphic
\( \omega ^{M} \) and antiholomorphic \( \hat{\omega }^{M} \)) is
conserved. This can be achieved by adding a term proportional to \( (P-\Pi )^{2} \)
to the action%
\footnote{Varying with respect to \( P_{z}^{m} \) and \( P^{m}_{\bar{z}} \)
yields for \( c\neq 0 \) algebraic equations of motion \( P_{z}^{m}=\Pi _{z}^{m} \)
and \( P^{m}_{\bar{z}}=\Pi ^{m}_{\bar{z}} \). Reinserting those equations
returns the old action. For \( c=1 \) the \( \Pi ^{2} \)-term cancels
with that of the original action and one gets a first order action.\( \quad \fussend  \) 
}\begin{eqnarray}
\mc {\tilde{L}} & = & \mc {L}+\frac{c}{2}\left( P_{z}^{m}-\Pi ^{m}_{z}\right) \left( P_{\bar{z}m}-\Pi _{\bar{z}m}\right) 
\end{eqnarray}
with some parameter \( c \) yet to be determined by the invariance
condition. All the other transformations remain unchanged%
\footnote{\label{keinp}The change of the transformation of \( d \) and \( \hat{d} \)
has to be implemented by appropriate changes in the transformation
of \( p \) and \( \hat{p} \). However, the information of \( \delta p \)
is completely encoded in the transformation of the other elementary
fields and the transformation of \( d \). In the following, we will
therefore only write down transformations of \( d \) instead of \( p \).\( \quad \fussend  \)
}. The variation of the action under local transformations reads:\begin{eqnarray}
\tilde{\delta }\tilde{S} & = & \int \tilde{\delta }\mc{L}+\frac{c}{2}\cdot \tilde{\delta }\left\{ \left( P_{z}^{m}-\Pi ^{m}_{z}\right) \left( P_{\bar{z}m}-\Pi _{\bar{z}m}\right) \right\} =\\
 & = & \int \delta \mc{L}+(\tilde{\delta }-\delta )d_{z\alpha }\, \bar{\partial }\theta ^{\alpha }+(\tilde{\delta }-\delta )\hat{d}_{\bar{z}\hat{\alpha }}\partial \hat{\theta }^{\hat{\alpha }}+ \\\nonumber
 &  & +c\cdot \left\{ \left( (\hat{\omega }\gamma ^{m}\partial \hat{\theta })-\frac{1}{2}\partial \hat{\omega }^{m}\right) \left( \Pi _{\bar{z}m}-P_{\bar{z}m}\right) +\left( (\omega \gamma ^{m}\bar{\partial }\theta )-\frac{1}{2}\bar{\partial }\omega ^{m}\right) \left( \Pi _{zm}-P_{zm}\right) \right\} =\\
 & = & \int \delta \mc{L}+\frac{c}{2}\bar{\partial }\omega ^{m}\left( P_{zm}-\Pi _{zm}\right) +\frac{c}{2}\partial \hat{\omega }^{m}\left( P_{\bar{z}m}-\Pi _{\bar{z}m}\right) + \\\nonumber
 &  & +\left( -2+c\right) (\omega \gamma ^{m}\bar{\partial }\theta )\left( \Pi _{zm}-P_{zm}\right) +\left( -2+c\right) (\hat{\omega }\gamma ^{m}\partial \hat{\theta })\left( \Pi _{\bar{z}m}-P_{\bar{z}m}\right)\label{localVariation} 
\end{eqnarray}
For the global variation to vanish, we thus have to choose \begin{eqnarray}
c & = & 2
\end{eqnarray}
The local variation shows that the old currents \( J_{m} \) and \( \hat{J}_{m} \)
change by \( \left( P_{mz}-\Pi _{zm}\right)  \) and \( \left( P_{\bar{z}m}-\Pi _{\bar{z}m}\right)  \)
respectively.\\
The action is now of the form (we drop the tilde \( \widetilde{\: } \)
of the new action and transformations)\begin{eqnarray} 	\hspace*{-9mm}
S & = & \int P^{m}_{z}P_{\bar{z}m}-P^{m}_{z}\Pi _{\bar{z}m}-\Pi ^{m}_{z}P_{\bar{z}m}+\frac{1}{2}\Pi ^{m}_{z}\Pi _{\bar{z}m}+\mc {L}_{WZ}+d_{z\alpha }\bar{\partial }\theta ^{\alpha }+\hat{d}_{\bar{z}\hat{\alpha }}\partial \hat{\theta }^{\hat{\alpha }}=\\					\hspace*{-9mm}
 & = & \int \left( P_{z}^{m}-\Pi ^{m}_{z}\right) \left( P_{\bar{z}m}-\Pi _{\bar{z}m}\right) -\frac{1}{2}\partial x^{m}\bar{\partial }x_{m}+p_{z\alpha }\bar{\partial }\theta ^{\alpha }+\hat{p}_{\bar{z}\hat{\alpha }}\bar{\partial }\hat{\theta }^{\hat{\alpha }}
\end{eqnarray}
and is invariant under global chiral and antichiral transformations.
The new local transformations read: \begin{eqnarray}
\delta d_{z\alpha } & = & i\partial \omega _{\alpha }+2(\gamma _{m}\omega )_{\alpha }P_{z}^{m}-2i\omega ^{m}(\gamma _{m}\partial \theta )_{\alpha }\label{dmitPtrafo} \\
\delta \hat{d}_{\bar{z}\hat{\alpha }} & = & i\bar{\partial }\hat{\omega }_{\hat{\alpha }}+2(\gamma _{m}\hat{\omega })_{\hat{\alpha }}P_{\bar{z}}^{m}-2i\hat{\omega }^{m}(\gamma _{m}\bar{\partial }\hat{\theta })_{\hat{\alpha }}\\
\delta P_{z}^{m} & = & -\partial \omega ^{m}+2(\omega \gamma ^{m}\partial \theta )\\
\delta P_{\bar{z}}^{m} & = & -\bar{\partial }\hat{\omega }^{m}+2(\hat{\omega }\gamma ^{m}\bar{\partial }\hat{\theta })\label{Ptrafo} 
\end{eqnarray}
The off-shell algebra for the local transformations is now closed.

\subsection{Gauging}

Gauging an algebra means introducing some gauge fields (connections)
such that the corresponding symmetry becomes local. Therefore one
gets a gauge symmetry with its generators (the currents) constrained
to zero. This can be done by an iterative procedure known under the
name Noether's method (see e.g. \cite{Deser:1970wk,VanNieuwenhuizen:ae}). \\
From the local variation (\ref{localVariation}) we have seen that the only currents
that change off-shell 
are \( \Pi _{zm}\To P_{zm} \) and \( \Pi _{\bar{z}m}\To P_{\bar{z}m} \).
The new currents are thus\begin{eqnarray}
J_{zM} & = & (P_{zm},id_{z\alpha },\partial \theta ^{\alpha })\\
\hat{J}_{\bar{z}\hat M} & = & (P_{\bar{z}m},i\hat{d}_{\bar{z}\hat{\alpha }},\bar{\partial }\hat{\theta }^{\hat{\alpha }})
\end{eqnarray}
They transform in the following way\begin{eqnarray}
\delta J_{M} & = & -\partial \omega _{M}+J_{P}f^{P}\tief{MN}\omega ^{N}\label{deltaJ} \\
\delta \hat{J}_{\hat{M}} & = & -\bar{\partial }\hat{\omega }_{\hat M}+\hat{J}_{\hat{P}}\hat{f}^{\hat{P}}\tief{\hat{M}\hat{N}}\hat{\omega }^{\hat{N}}
\end{eqnarray}
Now we are going to perfom the \textbf{Noether procedure}. For simplicity
we will stay in the condensed notation and only consider the chiral
symmetry. The antichiral symmetry is treated analogously. We are going
to introduce gauge fields \( A_{\bar{z}}^{M} \) in order to make
the symmetries local. It is useful for bookkeeping to define a grading
that counts the number of gauge fields. This induces a grading for
the variation \( \delta  \), depending on whether it increases or
decreases the grading of the fields. We will denote the grading by
a lower index. So far we have\begin{eqnarray}		\hspace*{-9mm}
S_{0} & = & \int P^{m}_{z}P_{\bar{z}m}-P^{m}_{z}\Pi _{\bar{z}m}-\Pi ^{m}_{z}P_{\bar{z}m}+\frac{1}{2}\Pi ^{m}_{z}\Pi _{\bar{z}m}+\mc {L}_{WZ}+d_{z\alpha }\bar{\partial }\theta ^{\alpha }+\hat{d}_{\bar{z}\hat{\alpha }}\partial \hat{\theta }^{\hat{\alpha }}\\					\hspace*{-9mm}
\delta _{0}S_{0} & = & \int J_{zM}\bar{\partial }\omega ^{M}
\end{eqnarray}
The next well known step is to add a coupling of gauge fields to the
currents\begin{equation}
S_{1}=-\int J_{zM}A_{\bar{z}}^{M}
\end{equation}
Defining \begin{equation}
\delta _{-1}A_{\bar{z}}^{M}=\bar{\partial }\omega ^{M}
\end{equation}
the variation of order zero vanishes (\( \delta _{-1}J_{M}=0 \) as
\( J_{M} \) does not contain any gauge fields and thus cannot be
decreased in their number) \begin{equation}
\delta _{0}S_{0}+\delta _{-1}S_{1}=0
\end{equation}
Terms of order 1 in the grading are \( \delta _{1}S_{0} \), \( \delta _{0}S_{1} \)
and \( \delta _{-1}S_{2} \). We would be done, if \( \delta _{0}S_{1} \)vanished
on its own.\begin{eqnarray}
\delta _{0}S_{1} & = & \int -\delta _{0}J_{zM}\, A_{\bar{z}}^{M}-J_{zM}\delta _{0}A_{\bar{z}}^{M}=\\
 & \stackrel{\rem{(\ref {deltaJ})}}{=} & \int -\left( J_{zP}f^{P}\tief{MN}\omega ^{N}-\partial \omega _{M}\right) A_{\bar{z}}^{M}-J_{zP}\delta _{0}A_{\bar{z}}^{P}\label{delta0S1} 
\end{eqnarray}
We can only cancel the first term by defining \begin{eqnarray}
\delta _{0}A_{\bar{z}}^{P} & = & -f^{P}\tief{MN}\omega ^{N}A_{\bar{z}}^{M}\\
\dann \quad \delta _{0}S_{1} & = & \int \partial \omega _{M}A_{\bar{z}}^{M}
\end{eqnarray}
The terms \( -\partial \omega _{M} \) in the transformation (\ref{deltaJ})
of the currents keep the procedure from terminating after this first
step. These terms correspond to the double poles in the current algebra
(\ref{algebra}). One can now use a simple trick and introduce a number
of auxiliary fields that produce the same algebra with a different
sign for the double poles. Adding those currents to the original ones
makes the double poles vanish. The simplest way to do this, is
to \textbf{double the fields} (including \( P \)), and subtract from
the original Lagrangian the same Lagrangian in terms of the auxiliary
fields.%
\footnote{In \cite[p.8]{Nh:2003kq} this trick was used to make the BRST transformations
nilpotent.\( \quad \fussend  \)
} This is similar to the gauged WZNW model, where one arrives at an
analogous doubling after gauge fixing and an adequate parametrization
of the surviving gauge fields \cite{Schnitzer:1989au,Schnitzer:1990dk}.
The original variables in WZNW are embedding functions \( g \) into
a group manifold, and the auxiliary fields are usually called \( h \).
The coordinates of the free theory do not parametrize a group manifold
(since \( p_{z\alpha } \) is an elementary field). But because of
this similarity, we put an index \( h \) to our auxiliary variables
and refer to the resulting currents as \( h \)-currents.\begin{eqnarray}
S & \equiv  & S_{\textrm{old}}-S_{h}
\end{eqnarray}
The \( h \)-action is of course seperately invariant under the same
chiral transformations expressed in \( h \)-coordinates as the original
Lagrangian. Let us call the corresponding transformation parameters
\( \omega _{M}^{h} \). For our purpose we have to choose \( \omega _{M}^{h}=\omega _{M} \).
In addition to (\ref{xtrafo}), (\ref{thetatrafo}), (\ref{thetahattrafo})
and (\ref{dmitPtrafo})-(\ref{Ptrafo}) we get the following transformations: \begin{eqnarray}
\delta x^{h\, m} & = & -\omega ^{m}+(\omega \gamma ^{m}\theta ^{h})-\hat{\omega }^{m}+(\hat{\omega }\gamma ^{m}\theta ^{h})\label{xhtrafo} \\
\delta \theta ^{h\, \alpha } & = & i\omega ^{\alpha }\\
\delta d^{h}_{z\alpha } & = & i\partial \omega _{\alpha }+2(\gamma _{m}\omega )_{\alpha }P_{z}^{m}-2i\omega ^{m}(\gamma _{m}\partial \theta )_{\alpha }\\
\delta P_{z}^{h\, m} & = & -\partial \omega ^{m}+2(\omega \gamma ^{m}\partial \theta )\\
\delta \hat{\theta }^{h\, \hat{\alpha }} & = & i\hat{\omega }^{\hat{\alpha }}\\
\delta \hat{d}^{h}_{\bar{z}\hat{\alpha }} & = & i\partial \hat{\omega }_{\hat{\alpha }}+2(\gamma _{m}\hat{\omega })_{\alpha }P_{\bar{z}}^{h\, m}-2i\hat{\omega }^{m}(\gamma _{m}\partial \hat{\theta }^{h})_{\hat{\alpha }}\\
\delta P_{\bar{z}}^{h\, m} & = & -\bar{\partial }\hat{\omega }^{m}+2(\hat{\omega }\gamma ^{m}\bar{\partial }\hat{\theta }^{h})\label{Phbarztrafo} 
\end{eqnarray}
Variation of the action under local transformations yields\begin{eqnarray}
\delta _{\omega }S & = & \int \left( J_{M}+J_{M}^{h}\right) \bar{\partial }\omega ^{M}+\left( \hat{J}_{\hat{M}}+\hat{J}_{\hat{M}}^{h}\right) \partial \hat{\omega }^{\hat{M}}\\
\textrm{where }J^{h}_{M} & \equiv  & -(P^{h}_{zm},id^{h}_{z\alpha },\partial \theta ^{h\alpha })\quad \hat{J}^{h}_{\hat{M}}\; \equiv \; -(P^{h}_{\bar{z}m},i\hat{d}^{h}_{\bar{z}\hat\alpha },\bar{\partial }\theta ^{h\hat{\alpha }})
\end{eqnarray}
 The transformation of the complete currents now has vanishing double
poles (no \( \partial \omega ^{M} \)-terms)\begin{eqnarray}
\delta (J_{zM}+J_{zM}^{h}) & = & (J_{zP}+J_{zP}^{h})f^{P}\tief{MN}\omega ^{N}\\
\delta (\hat{J}_{\bar{z}\hat{M}}+\hat{J}^{h}_{\bar{z}\hat{M}}) & = & (\hat{J}_{\bar{z}P}+\hat{J}^{h}_{\bar{z}P})\hat{f}^{\hat{P}}\tief{\hat{M}\hat{N}}\hat{\omega} ^{\hat{N}}
\end{eqnarray}
and the gauged action only needs the coupling to the connection\begin{equation}
\label{Sgauged}
S_{\textrm{gauged}}=S-S_{h}-\int (J_{M}+J^{h}_{M})A_{\bar{z}}^{M}-\int (\hat{J}_{\hat{M}}+\hat{J}^{h}_{\hat{M}})\hat{A}_{z}^{\hat{M}}
\end{equation}
with \begin{eqnarray}
\delta A_{\bar{z}}^{P} & = & \bar{\partial }\omega ^{P}-f^{P}\tief{MN}\omega ^{N}A_{\bar{z}}^{M}\\
\delta \hat{A}_{z}^{\hat{P}} & = & \partial \hat{\omega }^{\hat{P}}-\hat{f}^{\hat{P}}\tief{\hat{M}\hat{N}}\hat{\omega }^{\hat{N}}\hat{A}_{z}^{\hat{M}}
\end{eqnarray}

\subsection{Gauge Fixing and BRST Transformation}

One can use the gauge freedom to put all the gauge fields to zero
again. Within the standard BRST formalism we introduce ghosts by
rewriting the transformation parameters as\begin{equation}
\omega ^{M}\equiv \Lambda c^{M},\quad \hat{\omega }^{\hat{M}}\equiv \Lambda \hat{c}^{\hat{M}}
\end{equation}
 with an anticommuting imaginary global parameter \( \Lambda  \)
and the ghost fields \( c^{M} \). The BRST differential \( s \)
on the elementary fields is then defined as the transformation without
this parameter\begin{equation}
\delta _{\omega }\phi ^{M}\equiv \Lambda \cdot s\phi ^{M}
\end{equation}
The action is of course still invariant under this transformation.
We add the usual gauge fixing and ghost term to the Lagrangian: \begin{eqnarray}
\mc {L}_{\textrm{qu}} & = & \mc {L}_{\textrm{gauged }}+s(b_{M}A_{\bar{z}}^{M})
\end{eqnarray}
with\begin{eqnarray}
sb_{M} & = & \Lambda _{M}\\
s\Lambda _{M} & = & 0
\end{eqnarray}
where \( \Lambda _{M} \) is the Lagrange multiplier field and has
nothing to do with the scalar parameter \( \Lambda  \). The BRST-transformation
of the ghosts is defined such that \( s \) becomes nilpotent.\\
From the gauge transformation of the gauge field \begin{eqnarray}
\delta A_{\bar{z}}^{P} & = & \bar{\partial }\omega ^{P}-f^{P}\tief{MN}\omega ^{N}A_{\bar{z}}^{M}
\end{eqnarray}
we can read off its BRST transformation by pulling the parameter \( \Lambda  \)
to the front:%
\footnote{The ghosts as well as the antighosts have an anticommuting body, i.e.
the grading of \( c^{M} \) is \[
\abs{c^{M}}=\abs{M}+1\]
It thus makes sense to have contractions of the form \( (-)^{P}b_{P}c^{P}\qquad \fussend  \)
}\begin{eqnarray}
sA_{\bar{z}}^{P} & = & \bar{\partial }c^{P}-(-)^{N+M+P}f^{P}\tief{MN}c^{N}A_{\bar{z}}^{M}
\\[1mm]
\dann \quad s(b_{P}A_{\bar{z}}^{P}) & = & \Lambda _{P}A_{\bar{z}}^{P}-(-)^{P}b_{P}\bar{\partial }c^{P}+(-)^{N+M}b_{P}f^{P}\tief{MN}c^{N}A_{\bar{z}}^{M}
\end{eqnarray}
The complete quantum action is thus \begin{eqnarray}
S_{\textrm{qu}} & = & S_{\textrm{gauged}}+\int \Lambda _{P}A_{\bar{z}}^{P}-(-)^{P}b_{P}\bar{\partial }c^{P}+(-)^{N+M}b_{P}f^{P}\tief{MN}c^{N}A_{\bar{z}}^{M}
\end{eqnarray}
Varying with respect to \( \Lambda _{M} \) and \( A_{\bar{z}}^{M} \)
yields algebraic equations for those fields\begin{eqnarray}
A_{\bar{z}}^{M} & = & 0\\
\Lambda _{M} & = & (J_{M}+J_{M}^{h})-(-)^{N+M}b_{P}f^{P}\tief{MN}c^{N}\label{LautrupOnshell} 
\end{eqnarray}
which can be reinserted to finally arrive at the following action
(which we will call simply \( S \) again): \begin{eqnarray}
S & = & \int P^{m}_{z}P_{\bar{z}m}-P^{m}_{z}\Pi _{\bar{z}m}-\Pi ^{m}_{z}P_{\bar{z}m}+\frac{1}{2}\Pi ^{m}_{z}\Pi _{\bar{z}m}+\mc {L}_{WZ}+d_{z\alpha }\bar{\partial }\theta ^{\alpha }+\hat{d}_{\bar{z}\hat{\alpha }}\partial \hat{\theta }^{\hat{\alpha }}+\nonumber \\
 &  & -\Big(P^{h\, m}_{z}P^{h}_{\bar{z}m}-P^{h\, m}_{z}\Pi ^{h}_{\bar{z}m}-\Pi ^{h\, m}_{z}P^{h}_{\bar{z}m}+\frac{1}{2}\Pi ^{h\, m}_{z}\Pi ^{h}_{\bar{z}m}+\mc {L}^{h}_{WZ}+d^{h}_{z\alpha }\bar{\partial }\theta ^{h\, \alpha }+\hat{d}^{h}_{\bar{z}\hat{\alpha }}\partial \hat{\theta }^{h\, \hat{\alpha }}\Big)+\nonumber \\
 &  & +\beta _{zm}\bar{\partial }\xi ^{m}+\omega _{z\alpha }\bar{\partial }\lambda ^{\alpha }+\kappa _{z}^{\alpha }\bar{\partial }\chi _{\alpha }+\hat{\beta }_{\bar{z}m}\partial \hat{\xi }^{m}+\hat{\omega }_{\bar{z}\hat{\alpha }}\partial\hat{\lambda }^{\hat{\alpha }}+\hat{\kappa }_{\bar{z}}^{\hat{\alpha }}\partial\hat{\chi }_{\hat{\alpha }}
\end{eqnarray}
 Here we have defined%
\footnote{Like for \( p_{z\alpha } \) we use a sign for the antighosts in the
action that differs from that in \cite{Nh:2003kq}.\( \quad \fussend  \)
} \begin{eqnarray}
c^{M} & \equiv  & (-\xi ^{m},\lambda ^{\alpha },\chi _{\alpha })\\
\dann \quad c_{M} & = & (-\xi _{m},i\chi _{\alpha },-i\lambda ^{\alpha })\\
\textrm{and}\quad b_{M} & \equiv  & (\beta _{zm},\omega _{z\alpha },\kappa _{z}^{\alpha })\\
\dann \quad b^{M} & = & (\beta _{z}^{m},i\kappa _{z}^{\alpha },-i\omega _{z\alpha })
\end{eqnarray}
and equivalent relations for the hatted ghosts. From the former
gauge transormations (\ref{xtrafo}), (\ref{thetatrafo}), (\ref{thetahattrafo}),
(\ref{dmitPtrafo})-(\ref{Ptrafo}) and (\ref{xhtrafo})-(\ref{Phbarztrafo})
we can read off the BRST transformations\begin{eqnarray}
sx^{m} & = & \xi ^{m}+(\lambda \gamma ^{m}\theta )+\hat{\xi }^{m}+(\hat{\lambda }\gamma ^{m}\hat{\theta })\\
s\theta ^{\alpha } & = & i\lambda ^{\alpha }\\
sd_{z\alpha } & = & -\partial \chi _{\alpha }+2(\gamma _{m}\lambda )_{\alpha }P^{m}_{z}+2i\xi ^{m}(\gamma _{m}\partial \theta )_{\alpha }\\
sP_{z}^{m} & = & \partial \xi ^{m}+2(\lambda \gamma ^{m}\partial \theta )\\
s\hat{\theta }^{\hat{\alpha }} & = & i\hat{\lambda }^{\hat{\alpha }}\\
s\hat{d}_{\bar{z}\hat{\alpha }} & = & -\bar{\partial }\hat{\chi }_{\hat{\alpha }}+2(\gamma _{m}\hat{\lambda })_{\hat{\alpha }}P_{\bar{z}}^{m}+2i\hat{\xi }^{m}(\gamma _{m}\bar{\partial }\hat{\theta })_{\hat{\alpha }}\\
sP_{\bar{z}}^{m} & = & \bar{\partial }\hat{\xi }^{m}+2(\hat{\lambda }\gamma ^{m}\bar{\partial }\hat{\theta })\\
sx^{h\, m} & = & \xi ^{m}+(\lambda \gamma ^{m}\theta ^{h})+\hat{\xi }^{m}+(\hat{\lambda }\gamma ^{m}\hat{\theta }^{h})\\
 & \ddots  & \nonumber 
\end{eqnarray}
The transformation of the supersymmetric momentum has the form\begin{eqnarray}
s\Pi ^{m}_{\mu} & = & \partial_\mu \xi ^{m}+2(\lambda \gamma ^{m}\partial_\mu \theta )+\partial_\mu\hat{\xi }^{m}+2(\hat{\lambda }\gamma ^{m}\partial_\mu\hat{\theta })
\end{eqnarray}
 Up to the signs of \( p_{z\alpha } \), \( d_{z\alpha } \), \( \beta_{zm} \),
\( \omega _{z\alpha } \) and \( \kappa _{z}^{\alpha } \) these transformations
coincide in the chiral sector after integrating out \( P_{zm} \)
and \( P_{\bar{z}m} \) with those in \cite{Nh:2003kq}.\\
In order to make the BRST transformations nilpotent, the ghosts have
to transform in the following way:%
\footnote{\label{sJN}Taking into accout the appropriate grading signs modifies
the general form of the gauge transformations to the following BRST
transformations of the currents\begin{eqnarray*}
sJ_{N} & = & (-)^{N+M}J_{P}f^{P}\tief{NM}c^{M}-\partial c_{N}
\end{eqnarray*}
Demanding nilpotency for this formulae in condensed notation yields
(using the Jacobi-identity) the transformation of the ghosts.\( \qquad \fussend  \)
}\begin{eqnarray}
sc^{M} & = & -(-)^{K}\frac{1}{2}f^{M}\tief{LK}c^{K}c^{L}\label{scM} 
\end{eqnarray}
or in detail\begin{eqnarray}
s\xi ^{m} & = & -i(\lambda \gamma ^{m}\lambda )\\
s\lambda ^{\alpha } & = & 0\\
s\chi _{\alpha } & = & 2(\gamma _{m}\lambda )_{\alpha }\xi ^{m}
\end{eqnarray}
and similar transformations for the hatted ghosts. The transformation
of the antighosts \( sb_{M}=\Lambda _{M} \) turns via (\ref{LautrupOnshell})
into\begin{eqnarray}
sb_{M} & = & (J_{M}+J_{M}^{h})-(-)^{N+M}b_{P}f^{P}\tief{MN}c^{N}\label{sbM} 
\end{eqnarray}
or in detail\begin{eqnarray}
s\beta _{zm} & = & (P_{zm}-P^{h}_{zm})-2(\kappa _{z}\gamma _{m}\lambda )\\
s\omega _{z\alpha } & = & (id_{z\alpha }-id^{h}_{z\alpha })-2i\beta _{zm}(\gamma ^{m}\lambda )_{\alpha }-2(\gamma _{m}\kappa _{z})_{\alpha }\xi ^{m}\\
s\kappa _{z}^{\alpha } & = & (\partial \theta ^{\alpha }-\partial \theta ^{h\, \alpha })
\end{eqnarray}

\subsection{BRST Current, Composite B-field and Energy Momentum Tensor}

The BRST current can be derived by making the anticommuting transformation
parameter \( \Lambda  \) of the BRST transformation \( \delta _{\Lambda }(\ldots )=\Lambda s(\ldots ) \)
local. From the chiral transformations we know that the variation
of the {}``matter part'' (without ghosts) is \( \int (J_{M}+J^{h}_{M})\bar{\partial }(\Lambda c^{M}) \).
The variation of the complete action under local BRST is thus
{\small
\begin{eqnarray}
  \delta _{\Lambda }S & = & \int (J_{M}+J^{h}_{M})\bar{\partial }
  (\Lambda c^{M})+(\hat{J}_{\hat{M}}+\hat{J}^{h}_{\hat{M}})\partial 
  (\Lambda \hat{c}^{\hat{M}})-\delta _{\Lambda }\left( (-)^{M}b_{M}
  \bar{\partial }c^{M}+(-)^{\hat{M}}\hat{b}_{\hat{M}}\bar{\partial }
  \hat{c}^{\hat{M}}\right) =\qquad\qquad \\
  & = & \int \bar{\partial }\Lambda \left( (-)^{M}(J_{M}+J^{h}_{M})c^{M}+
  b_{M}sc^{M}\right) +\partial \Lambda 
  \left( (-)^{\hat{M}}(\hat{J}_{\hat{M}}+\hat{J}^{h}_{\hat{M}})
  \hat{c}^{\hat{M}}+\hat{b}_{\hat{M}}s\hat{c}^{\hat{M}}\right) 
\end{eqnarray}
}%
We can read off the \textbf{BRST currents}
{\small
\begin{eqnarray}
j^{B}_{z} & = & (-)^{M}(J_{M}+J^{h}_{M})c^{M}-(-)^{K}\frac{1}{2}b_{M}f^{M}\tief{LK}c^{K}c^{L}=\label{jcond} \\
 & = & -(P_{zm}-P^{h}_{zm})\xi ^{m}-(id_{z\alpha }-id^{h}_{z\alpha })\lambda ^{\alpha }-(\partial \theta ^{\alpha }-\partial \theta ^{h\, \alpha })\chi _{\alpha }+i\beta _{zm}(\lambda \gamma ^{m}\lambda )+2(\kappa _{z}\gamma _{m}\lambda )\xi ^{m}\qquad\qquad \\
\hat{\jmath }^{B}_{\bar{z}} & = & -(P_{\bar{z}m}-P^{h}_{\bar{z}m})\hat{\xi }^{m}-(i\hat{d}_{\bar{z}\hat{\alpha }}-i\hat{d}^{h}_{\bar{z}\hat{\alpha }})\hat{\lambda }^{\hat{\alpha }}-(\bar{\partial }\hat{\theta }^{\hat{\alpha }}-\bar{\partial }\hat{\theta }^{h\, \hat{\alpha }})\hat{\chi }_{\hat{\alpha }}+i\hat{\beta }_{\bar{z}m}(\hat{\lambda }\gamma ^{m}\hat{\lambda })+2(\hat{\kappa }_{\bar{z}}\gamma _{m}\hat{\lambda })\hat{\xi }^{m}
\end{eqnarray}
}%
which on-shell coincide with that of the chiral or antichiral heterotic
string respectively.

For gauged WZNW models there exists in general an operator \( B_{zz} \)
which makes \( T_{zz} \) BRST-exact 
(see \cite[p.24]{Figueroa-O'Farrill:1996cz})

\begin{eqnarray}
T_{zz} & = & \left[ Q,B_{zz}\right] \label{Texakt} \\
\textrm{where }Q & = & \oint j^{\textrm{B}}
\end{eqnarray}
For the covariant superstring based on the gauged WZNW model, it is
explicitely written down in \cite[p.11]{Nh:2003kq} and is important
to build up the \( N=2 \) superconformal algebra together with the
BRST-current, \( T_{zz} \) and the ghost current. Having carefully
performed the local BRST variation above makes it now simple for
us to recognize the symmetry corresponding to \( B_{zz} \) in these
calculations: Looking at the ghost action\begin{equation}
S_{gh}=\int -(-)^{M}b_{M}\bar{\partial }c^{M}-(-)^{\hat{M}}\hat{b}_{\hat{M}}\partial \hat{c}^{\hat{M}}=\int -(-)^{M}c_{M}\bar{\partial }b^{M}-(-)^{\hat{M}}\hat{c}_{\hat{M}}\partial\hat{b}^{\hat{M}}
\end{equation}
it becomes clear that \( b_{M} \) and \( c^{M} \) can interchange
their role, as long as the conformal weight of \( b_{M} \) is of
no importance. That means one can construct a new symmetry by interchanging
\( c^{M}\leftrightarrow b^{M} \) and \( \hat{c}^{\hat{M}}\leftrightarrow \hat{b}^{\hat{M}} \),
or in detail\begin{equation}
\begin{array}{ccccccc}
-\xi ^{m} & \leftrightarrow  & \beta _{z}^{m} &  & -\hat{\xi }^{m} & \leftrightarrow  & \hat{\beta }_{\bar{z}}^{m}\\
\lambda ^{\alpha } & \leftrightarrow  & i\kappa _{z}^{\alpha } & \textrm{and} & \hat{\lambda }^{\hat{\alpha }} & \leftrightarrow  & i\hat{\kappa }_{\bar{z}}^{\hat{\alpha }}\\
\chi _{\alpha } & \leftrightarrow  & -i\omega _{z\alpha } &  & \hat{\chi }_{\hat{\alpha }} & \leftrightarrow  & -i\hat{\omega }_{\bar{z}\hat{\alpha }}
\end{array}
\end{equation}
Performing this exchange in all BRST transformations would yield another
fermionic nilpotent transformation. However, the aim is that the generator
fulfills (\ref{Texakt}). \( T_{zz} \) is basically the square of
the original currents minus the square of the \( h \)-currents plus
ghost terms. In the BRST-current, we have \( (J+J^{h}) \)-terms.
We therefore need \( (J-J^{h}) \) terms in the \( B_{zz} \)-current.
Changing the relative sign of the transformation parameter for the
original fields and for the \( h \)-fields does not affect the invariance
of the matter action, as the \( h \)-part and the original part are
invariant independently. The resulting contribution to the current
from the matter part is then the difference of \( J_{M} \) and \( J_{M}^{h} \),
as desired. Call \( \Lambda _{B} \) the transformation parameter
corresponding to \( B_{zz} \) and \( t \) the fermionic transformation
without this parameter\begin{equation}
\delta _{\Lambda _{B}}(\ldots )\equiv \Lambda _{B}t(\ldots )
\end{equation}
 The variation of the complete action with respect to this transformation
yields\begin{eqnarray}
\delta _{\Lambda _{B}}S & = & \int \bar{\partial }\Lambda _{B}\left( (-)^{M}(J_{M}-J^{h}_{M})b^{M}+c_{M}tb^{M}\right) +(-)^{M}\Lambda _{B}(J_{M}-J^{h}_{M})\bar{\partial }b^{M}+\nonumber \\
 &  & -(-)^{M}\Lambda _{B}tc_{M}\bar{\partial }b^{M}+\Lambda _{B}c_{M}\bar{\partial }tb^{M}+\nonumber \\
 &  & +\partial \Lambda _{B}\left( (-)^{\hat{M}}(\hat{J}_{\hat{M}}-\hat{J}^{h}_{\hat{M}})\hat{b}^{\hat{M}}+\hat{c}_{\hat{M}}t\hat{b}^{\hat{M}}\right) +(-)^{\hat{M}}\Lambda _{B}(\hat{J}_{\hat{M}}-\hat{J}^{h}_{\hat{M}})\bar{\partial }\hat{b}^{\hat{M}}+\nonumber \\
 &  & -(-)^{\hat{M}}\Lambda _{B}t\hat{c}_{\hat{M}}\bar{\partial }\hat{b}^{\hat{M}}+\Lambda _{B}\hat{c}_{\hat{M}}\bar{\partial }t\hat{b}^{\hat{M}}
\end{eqnarray}
One can easily complete this transformation to a global symmetry of
the whole action by defining \( tc_{M}=(J_{M}-J^{h}_{M}) \) and \( tb^{M}=0 \)
and the same for the hatted ghosts. Nilpotency (which is the reason
not to make this simple choice in the case of BRST symmetry) is lost
already after changing the relative sign between the currents. The
holomorphic current for this new symmetry can then be read off to
be \( (-)^{M}(J_{M}-J^{h}_{M})b^{M} \). In order to obtain the proper
energy momentum tensor in an OPE with \( j^{B} \), this current
has to be multiplied with an additional factor of \( -\frac{1}{2} \):\begin{eqnarray}
B_{zz} & = & -\frac{1}{2}(P_{zm}+P^{h}_{zm})\beta_{z} ^{m}+\frac{i}{2}(id_{z\alpha }+id^{h}_{z\alpha })\kappa ^{\alpha }_{z}-\frac{i}{2}(\partial \theta ^{\alpha }+\partial \theta ^{h\, \alpha })\omega _{z\alpha }\\
B_{\bar{z}\bar{z}} & = & -\frac{1}{2}(P_{\bar{z}m}+P^{h}_{\bar{z}m})\hat{\beta }_{\bar{z}}^{m}+\frac{i}{2}(i\hat{d}_{\bar{z}\hat{\alpha }}+i\hat{d}^{h}_{\bar{z}\hat{\alpha }})\hat{\kappa }^{\hat{\alpha }}_{\bar{z}}-\frac{i}{2}(\bar{\partial }\hat{\theta }^{\hat{\alpha }}+\bar{\partial }\hat{\theta }^{h\, \hat{\alpha }})\hat{\omega }_{\bar{z}\hat{\alpha }}
\end{eqnarray}
The so defined \( B_{zz} \) is a homotopy for the \textbf{energy
momentum tensor} \( T_{zz} \) only on the operator level and not
as an off shell current:%
\footnote{Attention: For \( J_{N}-J_{N}^{h} \), the double poles do not vanish!\begin{eqnarray*}
s(J_{N}-J_{N}^{h}) & = & (-)^{N+M}(J_{P}-J_{P}^{h})f^{P}\tief{NM}c^{M}-2\partial c_{N}\quad \fussend 
\end{eqnarray*}

}\begin{eqnarray}
sB_{zz} & = & -(-)^{M}\frac{1}{2}\left( (J_{P}-J_{P}^{h})f^{P}\tief{NM}c^{M}-2\partial c_{N}\right) b^{N}+\nonumber \\
 &  & -(-)^{M}\frac{1}{2}(J^{M}-J^{h\, M})\left( (J_{M}+J_{M}^{h})-(-)^{N+M}b_{P}f^{P}\tief{MN}c^{N}\right) \\
 & = & -\frac{1}{2}J_{M}J^{M}+\frac{1}{2}J_{M}^{h}J^{h\, M}-(-)^{M}b_{M}\partial c^{M}=\\
 & = & -\frac{1}{2}P_{zm}P_{z}^{m}+d_{z\alpha }\partial \theta ^{\alpha }+\frac{1}{2}P^{h}_{zm}P_{z}^{h\, m}-d^{h}_{z\alpha }\partial \theta ^{h\, \alpha }+\nonumber\\
 &  & +\beta _{zm}\partial \xi ^{m}+\omega _{z\alpha }\partial \lambda ^{\alpha }+\kappa ^{\alpha }_{z}\partial \chi _{\alpha }=\\
 & \stackrel{\textrm{on shell}}{=} & T_{zz}
\end{eqnarray}
The off-shell holomorphic energy momentum tensor instead includes
\( \Pi _{z} \) and the hatted fields:\begin{eqnarray}
T_{zz} & = & \left( P_{z}-\Pi _{z}\right) ^{2}-\frac{1}{2}\Pi _{zm}\Pi _{z}^{m}+d_{z\alpha }\partial \theta ^{\alpha }+\hat{d}_{z\hat{\alpha }}\partial \hat{\theta }^{\hat{\alpha }}+\nonumber \\
 &  & -\left( P^{h}_{z}-\Pi ^{h}_{z}\right) ^{2}+\frac{1}{2}\Pi ^{h}_{zm}\Pi _{z}^{h\, m}-d^{h}_{z\alpha }\partial \theta ^{h\, \alpha }-\hat{d}^{h}_{z\hat{\alpha }}\partial \hat{\theta }^{h\, \hat{\alpha }}+\nonumber \\
 &  & +\beta _{zm}\partial \xi ^{m}+\omega _{z\alpha }\partial \lambda ^{\alpha }+\kappa ^{\alpha }_{z}\partial \chi _{\alpha }+\hat{\beta }_{zm}\partial \hat{\xi }^{m}+\hat{\omega }_{z\hat{\alpha }}\partial \hat{\lambda }^{\hat{\alpha }}+\hat{\kappa }^{\hat{\alpha }}_{z}\partial \hat{\chi }_{\hat{\alpha }}
\end{eqnarray}
 Similarly we have on shell \( sB_{\bar{z}\bar{z}}=T_{\bar{z}\bar{z}} \).

\section{The Second BRST Operator}

\label{secondBRST}Following the ideas of \cite{Nh:2004cz} we now
introduce a second BRST-operator and define a relative cohomology
whose purpose is to undo the gauging of \( \Pi _{z} \), 
\( \partial \theta  \),
\( \Pi _{\bar{z}} \) and \( \bar{\partial }\hat{\theta } \). We
review the essential ideas and extend the construction to an arbitrary
set of constraints that generate a first class system.

Consider a gauge algebra with generators \( G_{M}=\oint J_{M} \)
\begin{eqnarray}
\left[ G_{M},G_{N}\right]  & = & G_{K}f^{K}\tief{MN}
\end{eqnarray}
and assume that we only want to gauge symmetries which correspond
to some subset of generators \( G_{\alpha } \) that do not generate
a subalgebra. Call the remaining generators \( G_{\mf{a}} \) \begin{eqnarray}
G_{M} & \equiv  & (G_{\mf{a}},G_{\alpha })\label{GM} 
\end{eqnarray}
In order to gauge the generators \( G_{\alpha } \) one has to gauge
at first the complete algebra and end up at the usual BRST operator
\( Q=\oint j^{\textrm{B}} \) of the form (compare (\ref{jcond}))\begin{equation}
Q=\oint (-)^{M}J_{M}c^{M}-(-)^{K}\frac{1}{2}b_{M}f^{M}\tief{LK}c^{K}c^{L}
\end{equation}
 The indices of the ghosts and antighosts split in the same way as
those of the generators \( G_{M} \) in (\ref{GM}). In order to undo the gauging
of \( G_{\mf{a}} \), we set the corresponding ghosts \( c^{\mf{a}} \)
cohomologically to zero by making them exact.
However, \( c^{\mf{a}} \) cannot 
be exact with respect to \( Q \), because the BRST transformation
of \( c^{\mf{a}} \) is already fixed to something nonzero. Therefore
a second BRST operator \( Q_{c}=\oint j_{c} \) and 
some new fields
have to be introduced. The old ghosts \( c^{\mf{a}} \) and antighosts
\( b_{\mf{a}} \), as well as all the new fields will be removed from
cohomology via the following diagram: \begin{equation}
\label{diagram}
\begin{array}{cccccc}
b_{\mf{a}} &  &  &  &  & c^{\mf{a}}\\
 & \os {Q_{c}}{\searrow } &  &  & \os {Q_{c}}{\nearrow } & \\
\downarrow K &  & \antifeld _{\mf{a}} & \feld ^{\mf{a}} &  & \le {\uparrow -K}\\
 & \os {Q}{\nearrow } &  &  & \os {Q}{\searrow } & \\
b'_{\mf{a}} &  &  &  &  & c'^{\mf{a}}
\end{array}
\end{equation}
Here \( c'^{\mf{a}} \) and \( b'_{\mf{a}} \) are new ghosts and
antighosts with grading \( \abs{\mf{a}}+1 \), while \( \antifeld _{\mf{a}} \)
and \( \feld ^{\mf{a}} \) are fields with ghost number \( 0 \) and
grading \( \abs{\mf{a}} \). \( K \) will be explained below. The
contribution of the new fields to the Lagrangian is\begin{eqnarray}
\mc {L}' & = & -(-)^{\mf{a}}b'_{\mf{a}}\bar{\partial }c'^{\mf{a}}+\antifeld _{\mf{a}}\bar{\partial }\feld ^{\mf{a}}
\end{eqnarray}
To comply with
(\ref{diagram}) \( Q \) has to be extended
by a term \( (-)^{\mf{a}}\antifeld _{\mf{a}}c'^{\mf{a}} \).\begin{eqnarray}
Q & = & \oint (-)^{M}J_{M}c^{M}-(-)^{K}\frac{1}{2}b_{M}f^{M}\tief{LK}c^{K}c^{L}+(-)^{\mf{a}}\antifeld _{\mf{a}}c'^{\mf{a}}\\
\dann \quad sb_{\mf{a}}' & = & \antifeld _{\mf{a}}\label{sba'} \\
s\feld ^{\mf{a}} & = & c'^{\mf{a}}
\end{eqnarray}
 One can construct a suitable \( Q_{c} \) that anticommutes with
\( Q \) as the commutator of \( Q \) with a homotopy operator~\( K \)
\begin{eqnarray}
K & \equiv  & \oint k\, \equiv \, \oint (-)^{\mf{a}}b'_{\mf{a}}c^{\mf{a}}\\
\dann \quad \delta _{k}c'^{\mf{a}} & = & -c^{\mf{a}}\\
\delta _{k}b_{\mf{a}} & = & b'_{\mf{a}}
\end{eqnarray}
 \begin{eqnarray}
Q_{c}\equiv \oint j_{c} & \equiv  & \left[ Q,K\right] \\
 & = & \oint (-)^{\mf{a}}\antifeld _{\mf{a}}c^{\mf{a}}+(-)^{N}\frac{1}{2}b'_{\mf{a}}f^{\mf{a}}\tief{MN}c^{N}c^{M}
\end{eqnarray}
with corresponding transformations\begin{eqnarray}
s_{c}b_{M} & = & \underbrace{s\delta _{k}b_{M}}_{\le {\antifeld _{\mf{a}}\textrm{ for }M=\mf{a}}}+(-)^{N+M}\left( b'_{\mf{b}}f^{\mf{b}}\tief{MN}c^{N}\right) \label{scbM} \\
s_{c}\feld ^{\mf{a}} & = & c^{\mf{a}}\\
s_{c}c'^{\mf{a}} & = & (-)^{K}\frac{1}{2}f^{\mf{a}}\tief{LK}c^{K}c^{L}=-sc^{\mf{a}}
\end{eqnarray}
The Jacobi identity implies \begin{equation}
\label{JacobiImplies}
\left[ Q,Q_{c}\right] =0\quad \textrm{and}\quad \left[ Q_{c},Q_{c}\right] 
=\left[ Q,\left[ K,Q_{c}\right] \right] 
\end{equation}
It is thus sufficient to check \( \delta _{K}Q_{c}=0 \), which is
obviously the case, to obtain nilpotency of the second BRST operator\begin{equation}
\left[ Q_{c},Q_{c}\right] =0
\end{equation}
Equation (\ref{scbM}) differs a little bit from the diagram (\ref{diagram}).
That does not hurt, as \( \pi _{\mf{a}} \) according to (\ref{sba'})
is already BRST exact with respect to \( Q \). Physical observables
are then defined to lie in the relative cohomology of \( Q \) with
respect to \( Q_{c} \).

Part of \( Q \) turns out to be exact with respect to \( Q_{c} \).
If we define \begin{eqnarray}
\Xi  & \equiv  & (-)^{\mf{a}}b_{\mf{a}}c'^{\mf{a}}+\left( J_{\mf{a}}-(-)^{\gamma +\mf{a}}b_{\alpha }f^{\alpha }\tief{\mf{a}\gamma }c^{\gamma }\right) \feld ^{\mf{a}}
\end{eqnarray}
then \( Q \) can be written as\begin{eqnarray}	\hspace*{-14mm}
Q & = & \oint (-)^{\alpha }J_{\alpha }c^{\alpha }-(-)^{\beta }\frac{1}{2}b_{\gamma }f^{\gamma }\tief{\alpha \beta }c^{\beta }c^{\alpha }+s_c\Xi +\nonumber 
\\					\hspace*{-6mm}
 &  & -(-)^{\mf{b}}\frac{1}{2}b_{\gamma }f^{\gamma }\tief{\mf{a}\mf{b}}c^{\mf{b}}c^{\mf{a}}-(-)^{N}b'_{\mf{b}}f^{\mf{b}}\tief{\mf{a}N}c^{N}c'^{\mf{a}}-(-)^{\gamma +\mf{a}+N+\alpha }b'_{\mf{b}}f^{\mf{b}}\tief{\alpha N}c^{N}f^{\alpha }\tief{\mf{a}\gamma }c^{\gamma }\feld ^{\mf{a}}
\end{eqnarray}
For the gauging of the roots of a simple Lie algebra   
all structure constants in the second line vanish \cite{Nh:2004cz}.
For \textbf{superstrings}, on the other hand 
\begin{eqnarray}
Q & = & \oint (-)^{\alpha }J_{\alpha }c^{\alpha }	+s_c\,\Xi 
-(-)^{\alpha}b'_{\mf{b}}f^{\mf{b}}\tief{\mf{a}\alpha}c^{\alpha}c'^{\mf{a}}
\end{eqnarray}
 for the chiral sector. It may therefore be necessary to modify $Q$ or/and
$K$ in order to get the Berkovits BRST charge up to $Q_c$-exact terms.\\
Actually, for the string there are two algebras with currents $J+J^h$
and $\hat J+\hat J^h$, where
\begin{eqnarray}			\hspace*{-6mm}
 & \begin{array}{ccccccc}
J_{M} & = & (J_{m},J_{\alpha },J_{\q{\alpha }})=(\Pi _{zm},id_{z\alpha },\partial \theta ^{\alpha }) &  & \hat{J}_{\hat{M}} & = & (\hat{J}_{\hat{m}},\hat{J}_{\hat{\alpha }},\hat{J}_{\q{\hat{\alpha }}})=(\Pi _{\bar{z}m},i\hat{d}_{\bar{z}\hat{\alpha }},\bar{\partial }\hat{\theta }^{\hat{\alpha }})
\end{array} & \\			\hspace*{-6mm}
 & \begin{array}{ccccccc}
\; J_{\mf{a}} & = & (J_{m},J_{\q{\alpha }})=(\Pi _{zm},\partial \theta ^{\alpha })\qquad \qquad  &  & \; \hat{J}_{\mf{\hat{a}}} & = & (\hat{J}_{\hat{m}},\hat{J}_{\q{\hat{\alpha }}})=(\Pi _{\bar{z}m},\bar{\partial }\hat{\theta }^{\hat{\alpha }})\qquad \qquad 
\end{array} & 
\end{eqnarray}
The detailed translation of the ghosts and antighosts and the new
fields looks as follows\begin{equation}
\begin{array}{ccccccc}
c^{\mf{a}} & \equiv  & (c^{m},c^{\q{\alpha }})=(-\xi ^{m},\chi _{\alpha }) &  & \hat{c}^{\mf{\hat{a}}} & \equiv  & (\hat{c}^{m},\hat{c}^{\q{\hat{\alpha }}})=(-\hat{\xi }^{m},\hat{\chi }_{\hat{\alpha }})\\
b_{\mf{a}} & \equiv  & (b_{m},c_{\q{\alpha }})=(\beta _{zm},\kappa^\alpha_z) &  & \hat{b}_{\mf{\hat{a}}} & \equiv  & (\hat{b}_{m},\hat{c}_{\q{\hat{\alpha }}})=(\hat{\beta }_{zm},\hat{\kappa }^{\hat\alpha}_{z})\\
c'^{\mf{a}} & \equiv  & (c'^{m},c'^{\q{\alpha }})=(-\xi '^{m},\chi '_{\alpha }) &  & \hat{c}'^{\mf{\hat{a}}} & \equiv  & (\hat{c}'^{m},\hat{c}'^{\q{\hat{\alpha }}})=(-\hat{\xi }'^{m},\hat{\chi }'_{\hat{\alpha }})\\
b'_{\mf{a}} & \equiv  & (b'_{m},c'_{\q{\alpha }})=(\beta '_{zm},\kappa '^{\alpha }_{z}) &  & \hat{b}'_{\mf{\hat{a}}} & \equiv  & (\hat{b}'_{m},\hat{c}'_{\q{\hat{\alpha }}})=(\hat{\beta }'_{zm},\hat{\kappa }'^{\hat{\alpha }}_{z})\\
\feld ^{\mf{a}} & \equiv  & (\feld ^{m},\feld ^{\q{\alpha }})\equiv (\feld ^{m},-i\feld _{\alpha }) &  & \hat{\antifeld }_{\mf{\hat{a}}} & \equiv  & (\hat{\antifeld }_{m},\hat{\antifeld }_{\q{\hat{\alpha }}})\equiv (\hat{\antifeld }_{m},-i\hat{\antifeld }^{\hat{\alpha }})\\
\antifeld _{\mf{a}} & \equiv  & (\antifeld _{m},\antifeld _{\q{\alpha }})\equiv (\antifeld _{m},-i\antifeld ^{\alpha }) &  & \hat{\antifeld }_{\mf{\hat{a}}} & \equiv  & (\hat{\antifeld }_{m},\hat{\antifeld }_{\q{\hat{\alpha }}})\equiv (\hat{\antifeld }_{m},-i\hat{\antifeld }^{\hat{\alpha }})
\end{array}
\end{equation}
 The general equations thus translate into\begin{eqnarray}
  \mc {L}' & = & \beta '_{zm}\bar{\partial }\xi '^{m}+\kappa _{z}'^{\alpha }\bar{\partial }\chi '_{\alpha }+\antifeld _{\mf{a}}\bar{\partial} \feld ^{\mf{a}}+\nonumber\\
& & +\hat{\beta }'_{\bar{z}m}\partial\hat{\xi }'^{m}+\hat{\kappa }_{\bar{z}}'^{\hat{\alpha }}\partial\hat{\chi }'_{\hat{\alpha }}+\hat{\antifeld }_{\mf{\hat{a}}}\partial \hat{\feld }^{\mf{\hat{a}}}
\end{eqnarray}
\begin{eqnarray}
j^{\textrm{B}}_{z} & = & -(P_{zm}-P^{h}_{zm})\xi ^{m}-(id_{z\alpha }-id^{h}_{z\alpha })\lambda ^{\alpha }-(\partial \theta ^{\alpha }-\partial \theta ^{h\, \alpha })\chi _{\alpha }+\nonumber \\
 &  & +i\beta _{zm}(\lambda \gamma ^{m}\lambda )+2(\kappa _{z}\gamma _{m}\lambda )\xi ^{m}-\antifeld _{m}\xi '^{m}-\antifeld _{\q{\alpha }}\chi _{\alpha }'\\
s\beta _{zm}' & = & \antifeld _{m}\\
s\kappa _{z}'^{\alpha } & = & \pi _{\q{\alpha }}=-i\pi ^{\alpha }\\
s\feld ^{m} & = & -\xi '^{m}\\
s\feld ^{\q{\alpha }} & = & -is\feld _{\alpha }=\chi '_{\alpha }
\end{eqnarray}
 \begin{eqnarray}
k_{z} & \equiv  & -\beta '_{zm}\xi ^{m}-\kappa _{z}'^{\alpha }\chi _{\alpha }\label{homotopy} \\
\delta _{k}\xi '^{m} & = & -\xi ^{m}\\
\delta _{k}\chi '_{\alpha } & = & -\chi _{\alpha }\\
\delta _{k}\beta _{zm} & = & \beta '_{zm}\\
\delta _{k}\kappa _{z}^{\alpha } & = & \kappa _{z}'^{\alpha }
\end{eqnarray}
\begin{eqnarray}
j_{cz} & \equiv  & sk_{z}=-\antifeld _{zm}\xi ^{m}-\antifeld _{z\q{\alpha }}\chi _{\alpha }-i\beta '_{zm}(\lambda \gamma ^{m}\lambda )-2(\kappa _{z}'\gamma _{m}\lambda )\xi ^{m}\\
s_{c}\beta _{zm} & = & \antifeld _{m}+2(\kappa'_{z}\gamma _{m}\lambda )\\
s_{c}\kappa _{z}^{\alpha } & = & \antifeld _{\q{\alpha }}\\
s_{c}\omega _{z\alpha } & = & 2i\beta'_{zm}(\gamma ^{m}\lambda )_{\alpha }+2(\gamma _{m}\kappa'_{z})_{\alpha }\xi ^{m}\\
s_{c}\feld ^{m} & = & -\xi ^{m}\\
s_{c}\feld ^{\q{\alpha }} & = & \chi _{\alpha }\\
s_{c}\xi '^{m} & = & i(\lambda \gamma ^{m}\lambda )=-s\xi ^{m}\\
s_{c}\chi '_{\alpha } & = & -2(\gamma _{m}\lambda )_{\alpha }\xi ^{m}=-s\chi _{\alpha }
\end{eqnarray}
and the same for the hatted fields, which are contained in the \( \bar{z} \)-component
of the currents.

The new fields contribute to the on-shell energy momentum tensor in
the following way\begin{eqnarray}
T_{zz} & \To  & T_{zz}-(-)^{\mf{a}}b'_{\mf{a}}\partial c'^{\mf{a}}+\antifeld _{\mf{a}}\partial \feld ^{\mf{a}}\\
 & = & T_{zz}+\beta '_{zm}\partial \xi '^{m}+\kappa '^{\alpha }_{z}\partial \chi '_{\alpha }+\antifeld _{z\mf{a}}\partial \feld ^{\mf{a}}
\end{eqnarray}
Changing the composite \( B \)-field \begin{eqnarray}
B_{zz} & \To  & B_{zz}+b'_{\mf{a}}\partial \feld ^{\mf{a}}
\end{eqnarray}
repairs the on-shell relation\begin{equation}
T_{zz}\stackrel{\textrm{on}-\textrm{shell}}{=}sB_{zz}
\end{equation}

\section{Operator Algebra}

\label{operatoralgebra}Left moving and rightmoving sector do not
mix on-shell, and we thus concentrate on the left-moving sector only.
On the operator level, we have the same algebra as in \cite{Nh:2003kq},
namely a Kazama algebra. The new ghosts and fields \( c'^{\mf{a}},b'_{\mf{a}},\feld ^{\mf{a}}\textrm{ and }\antifeld _{\mf{a}} \)
do not disturb this structure, but change the ghost number anomaly.
For completeness, we present the algebra here again:%
\footnote{We used the OPE package OPE-defs.m by K. Thielemans \cite{Thielemans:1992mu}.\( \quad \fussend  \)
}{\allowdisplaybreaks\begin{eqnarray}
T(z)T(w) & \sim  & \frac{2T(w)}{(z-w)^{2}}+\frac{\partial T(w)}{z-w}\\
T(z)j^{\textrm{B}}(w) & \sim  & \frac{j^{\textrm{B}}(w)}{(z-w)^{2}}+\frac{\partial j^{\textrm{B}}(w)}{z-w}\\
T(z)B(w) & \sim  & \frac{2B(w)}{(z-w)^{2}}+\frac{\partial B(w)}{z-w}\\
T(z)j^{\textrm{gh}}(w) & \sim  & \frac{28}{(z-w)^{3}}+\frac{j^{\textrm{gh}}(w)}{(z-w)^{2}}+\frac{\partial j^{\textrm{gh}}(w)}{z-w}=\frac{28}{(z-w)^{3}}+\frac{j^{\textrm{gh}}(z)}{(z-w)^{2}}\\
j^{\textrm{B}}(z)B(w) & \sim  & \frac{-28}{(z-w)^{3}}+\frac{j^{\textrm{gh}}(w)}{(z-w)^{2}}+\frac{T(w)}{z-w}\\
j^{\textrm{gh}}(z)j^{\textrm{B}}(w) & \sim  & \frac{j^{\textrm{B}}(w)}{z-w}\\
j^{\textrm{gh}}(z)B(w) & \sim  & \frac{-B(w)}{z-w}\\
j^{\textrm{gh}}(z)j^{\textrm{gh}}(w) & \sim  & \frac{-28}{(z-w)^{2}}\\
 &  & \nonumber \\
B(z)B(w) & \sim  & \frac{F(w)}{z-w}\\
j^{\textrm{B}}(z)\Phi (w) & \sim  & \frac{F(w)}{z-w}
\end{eqnarray}
with\begin{eqnarray}			\hspace*{-9mm}
T_{zz} & = & -\frac{1}{2}\Pi _{zm}\Pi _{z}^{m}+d_{z\alpha }\partial \theta ^{\alpha }+\frac{1}{2}\Pi ^{h}_{zm}\Pi _{z}^{h\, m}-d^{h}_{z\alpha }\partial \theta ^{h\, \alpha }+\nonumber \\		\hspace*{-9mm}
 &  & +\beta _{zm}\partial \xi ^{m}+\omega _{z\alpha }\partial \lambda ^{\alpha }+\kappa ^{\alpha }_{z}\partial \chi _{\alpha }+\beta '_{zm}\partial \xi '^{m}+\kappa '^{\alpha }_{z}\partial \chi '_{\alpha }+\antifeld _{z\mf{a}}\partial \feld ^{\mf{a}}\\			\hspace*{-9mm}
j_{z}^{\textrm{B}} & = & -(\Pi _{zm}-\Pi ^{h}_{zm})\xi ^{m}-(id_{z\alpha }-id^{h}_{z\alpha })\lambda ^{\alpha }-(\partial \theta ^{\alpha }-\partial \theta ^{h\, \alpha })\chi _{\alpha }+\nonumber \\	\hspace*{-9mm}
 &  & +i\beta _{zm}(\lambda \gamma ^{m}\lambda )+2(\kappa _{z}\gamma _{m}\lambda )\xi ^{m}-\antifeld _{m}\xi '^{m}-\antifeld _{\q{\alpha }}\chi _{\alpha }'
\\					\hspace*{-9mm}
B_{zz} & = & -\frac{1}{2}(\Pi _{zm}+\Pi ^{h}_{zm})\beta_{z}^{m}+\frac{i}{2}(id_{z\alpha }+id^{h}_{z\alpha })\kappa ^{\alpha }_{z}-\frac{i}{2}(\partial \theta ^{\alpha }+\partial \theta ^{h\, \alpha })\omega _{z\alpha }+b'_{\mf{a}}\partial \feld ^{\mf{a}}\\				\hspace*{-9mm}
j_{z}^{\textrm{gh}} & = & \beta _{zm}\xi ^{m}+\omega _{z\alpha }\lambda ^{\alpha }+\kappa _{z}^{\alpha }\chi _{\alpha }+\beta '_{zm}\xi '^{m}+\kappa _{z}'^{\alpha }\chi '_{\alpha }\\	\label{Fzzz}	\hspace*{-9mm}
F_{zzz} & \equiv  & -i\beta _{zm}\left( \kappa _{z}\gamma ^{m}(\partial \theta -\partial \theta ^{h})\right) +\frac{i}{2}(\kappa _{z}\gamma ^{m}\kappa _{z})\left( \Pi _{zm}-\Pi _{zm}^{h}\right) \\	\hspace*{-9mm}
\Phi _{zzz} & \equiv  & \frac{i}{2}\beta _{zm}(\kappa _{z}\gamma ^{m}\kappa _{z})\label{Phi}
\end{eqnarray}
}The only term that prevents the algebra from coinciding with an
\( N=2 \) superconformal algebra is the BRST-exact operator \( F \).
To turn the Kazama algebra into an \( N=2 \) superconformal algebra,
\( B \) can be made nilpotent by adding a quartet of two anticommuting
\( (b'_{zz},c'^{z}) \) and two commuting ghosts \( (\beta '_{zz},\gamma '^{z}) \)
as it is done in \cite{Nh:2003kq}%
\footnote{\label{falscherBRST}In \cite{Nh:2003kq} this quartet is called
\( (b_{zz},c^{z},\beta _{zz},\gamma ^{z}) \), and \( b_{zz} \) and
\( c^{z} \) are later identified with the worldsheet diffeomorphism
ghosts. However, we disagree with the final BRST charge \( \bf {Q} \)
of \cite{Nh:2003kq}, eqs. (6.1)-(6.3), because it is not nilpotent.
In particular, the operator \( Q_{V} \) defined in (6.1) does not square to zero. 
It might be that the resulting terms are \( Q_{c} \) exact, which
would be enough to build a relative cohomology, but we decided to
seperate the steps of making \( B_{zz} \) nilpotent via a first quartet
\( (b'_{zz},c'^{z},\beta '_{zz},\gamma '^{z}) \) and of including
worldsheet diffeomorphism invariance via a second quartet \( (b_{zz},c^{z},\beta _{zz},\gamma ^{z}) \).\( \qquad \fussend  \)
}. The ghosts form a quartet 
\begin{eqnarray}
sc'^{z} & \equiv  & -\gamma '^{z}\\
s\beta '_{zz} & \equiv  & -b'_{zz}
\end{eqnarray}
and thus do not contribute to the cohomology.
Having an additional term \( b'_{zz}\bar{\partial }c'^{z}+\beta '_{zz}\bar{\partial }\gamma '^{z} \)
in the Lagrangian, this corresponds to the following new term in the
BRST-current:\begin{equation}
j^{\textrm{B}}\To j^{\textrm{B}}+b'_{zz}\gamma '^{z}
\end{equation}
 The term \( F \) in the algebra disappears when one changes \( B \)
to \begin{equation}
B_{zz}\To B_{zz}-2\beta '_{zz}\partial c'^{z}-c'^{z}\partial \beta '_{zz}-b'_{zz}-\frac{1}{2}c'^{z}F_{zzz}-\frac{1}{2}\gamma '^{z}\Phi _{zzz}
\end{equation}
The energy momentum tensor now reads\begin{eqnarray}
T_{zz} & = & -\frac{1}{2}\Pi _{zm}\Pi _{z}^{m}+d_{z\alpha }\partial \theta ^{\alpha }+\frac{1}{2}\Pi ^{h}_{zm}\Pi _{z}^{h\, m}-d^{h}_{z\alpha }\partial \theta ^{h\, \alpha }+\beta _{zm}\partial \xi ^{m}+\omega _{z\alpha }\partial \lambda ^{\alpha }+\kappa ^{\alpha }_{z}\partial \chi _{\alpha }+\nonumber \\
 &  & +\beta '_{zm}\partial \xi '^{m}+\kappa '^{\alpha }_{z}\partial \chi '_{\alpha }+\antifeld _{z\mf{a}}\partial \feld ^{\mf{a}}+2\beta '_{zz}\partial \gamma '^{z}+\partial \beta '_{zz}\gamma '^{z}+2b'_{zz}\partial c'^{z}+\partial b'_{zz}c'^{z}
\end{eqnarray}
and the ghost current becomes\begin{equation}
j_{z}^{\textrm{gh}}=\beta _{zm}\xi ^{m}+\omega _{z\alpha }\lambda ^{\alpha }+\kappa _{z}^{\alpha }\chi _{\alpha }+\beta '_{zm}\xi '^{m}+\kappa _{z}'^{\alpha }\chi '_{\alpha }+b'_{zz}c'^{z}+2\beta '_{zz}\gamma '^{z}
\end{equation}
The resulting current algebra is a twisted \( N=2 \) superconformal
algebra, as was already shown in \cite{Nh:2003kq}:\begin{eqnarray}
T(z)T(w) & \sim  & \frac{2T(w)}{(z-w)^{2}}+\frac{\partial T(w)}{z-w}\\
T(z)j^{\textrm{B}}(w) & \sim  & \frac{j^{\textrm{B}}(w)}{(z-w)^{2}}+\frac{\partial j^{\textrm{B}}(w)}{z-w}\\
T(z)B(w) & \sim  & \frac{2B(w)}{(z-w)^{2}}+\frac{\partial B(w)}{z-w}\\
T(z)j^{\textrm{gh}}(w) & \sim  & \frac{31}{(z-w)^{3}}+\frac{j^{\textrm{gh}}(w)}{(z-w)^{2}}+\frac{\partial j^{\textrm{gh}}(w)}{z-w}=\frac{31}{(z-w)^{3}}+\frac{j^{\textrm{gh}}(z)}{(z-w)^{2}}\\
j^{\textrm{B}}(z)B(w) & \sim  & \frac{-31}{(z-w)^{3}}+\frac{j^{\textrm{gh}}(w)}{(z-w)^{2}}+\frac{T(w)}{z-w}\label{jB} \\
j^{\textrm{gh}}(z)j^{\textrm{B}}(w) & \sim  & \frac{j^{\textrm{B}}(w)}{z-w}\\
j^{\textrm{gh}}(z)B(w) & \sim  & \frac{-B(w)}{z-w}\\
j^{\textrm{gh}}(z)j^{\textrm{gh}}(w) & \sim  & \frac{-31}{(z-w)^{2}}\\
 &  & \nonumber \\
B(z)B(w) & \sim  & 0
\end{eqnarray}
The usual structure can be seen by defining operators \( \breve{T}\equiv T-\frac{1}{2}\partial j^{\textrm{gh}},\: J\equiv j^{\textrm{gh}},\: G^{+}\equiv j^{\textrm{B}},\: G^{-}\equiv 2B \).
This is basically a topological twist in \( T \). The central charge
of the superconformal algebra becomes \( c=-93 \):\begin{eqnarray}
\breve{T}(z)\breve{T}(w) & \sim  & \frac{c/2}{(z-w)^{4}}+\frac{2\breve{T}(w)}{(z-w)^{2}}+\frac{\partial \breve{T}(w)}{z-w}\\
\breve{T}(z)G^{\pm } & \sim  & \frac{3/2\cdot G^{\pm }(w)}{(z-w)^{2}}+\frac{\partial G^{\pm }(w)}{z-w}\\
\breve{T}(z)J(w) & \sim  & \frac{J(w)}{(z-w)^{2}}+\frac{\partial J(w)}{z-w}=\frac{J(z)}{(z-w)^{2}}\\
G^{+}(z)G^{-}(w) & \sim  & \frac{2c/3}{(z-w)^{3}}+\frac{2J(w)}{(z-w)^{2}}+\frac{2\breve{T}(w)+\partial J(w)}{z-w}\\
J(z)G^{\pm }(w) & \sim  & \frac{\pm J(w)}{z-w}\\
J(z)J(w) & \sim  & \frac{c/3}{(z-w)^{2}}
\end{eqnarray}

\section{Worldsheet Diffeomorphism Invariance}

\label{diffeomorphismus}In order to implement worldsheet diffeomorphism
invariance, we would have to gauge the symmetry corresponding to \( T_{zz} \).
The algebra (especially (\ref{jB})) tells us that we then also have
to gauge \( B_{zz} \)%
\footnote{In the twisted theory, this would mean gauging one of the supersymmetry
generators.\( \quad \fussend  \)
}. We have not explicitely performed this gauging, but after gauge
fixing, one expects the additional terms \( c^{z}(T_{zz}+\frac{1}{2}T^{\textrm{top}}_{zz})+\gamma ^{z}(B_{zz}+\frac{1}{2}B^{\textrm{top}}_{zz}) \)
as given in \cite[p.125]{Dijkgraaf:1990qw}. This was of course the
idea in \cite{Nh:2003kq}, but there the final BRST operator
was not nilpotent as mentioned in footnote \ref{falscherBRST}. We
now add the topological quartet \( (b_{zz},c^{z},\beta _{zz},\gamma ^{z}) \).
This quartet itself obeys an \( N=2 \) superconformal algebra, if
one defines \begin{eqnarray}
T^{\textrm{top}}_{zz} & = & 2\beta _{zz}\partial \gamma ^{z}+\partial \beta _{zz}\gamma ^{z}+2b_{zz}\partial c^{z}+\partial b_{zz}c^{z}\\
j_{z}^{\textrm{B},\textrm{top}} & = & b_{zz}\gamma ^{z}\\
B_{zz}^{\textrm{top}} & = & -2\partial c^{z}\, \beta _{zz}-c^{z}\partial \beta _{zz}\\
j^{\textrm{gh},\textrm{top}} & = & b_{zz}c^{z}+2\beta _{zz}\gamma ^{z}
\end{eqnarray}
The ghosts decouple from the cohomology because \begin{eqnarray}
sc^{z} & \equiv  & -\gamma ^{z}\\
s\beta _{zz} & \equiv  & -b_{zz}
\end{eqnarray}
And the Lagrangian is of the form \begin{equation}
\mc {L}^{\textrm{top}}=b_{zz}\bar{\partial }c^{z}+\beta _{zz}\bar{\partial }\gamma ^{z}
\end{equation}
 We want to keep \( T \) BRST-exact, and therefore we need \( j_{z}^{\textrm{B},\textrm{top}} \)
as part of our final BRST current:\begin{eqnarray}
T_{zz} & = & \left[ Q,B_{zz}\right] \\
T_{zz}^{\textrm{top}} & = & \left[ Q^{\textrm{top}},B^{\textrm{top}}_{zz}\right] 
\end{eqnarray}
 \bmcI\begin{equation}
\dann \quad \left[ Q+Q^{\textrm{top}}+\oint c^{z}\Big (T_{zz}+\frac{1}{2}T_{zz}^{\textrm{top}}\Big )+\gamma ^{z}\Big (B_{zz}+\frac{1}{2}B_{zz}^{\textrm{top}}\Big)\: ,\: B_{zz}+B_{zz}^{\textrm{top}}\right] =T_{zz}+T_{zz}^{\textrm{top}}
\end{equation}
\emcI The final BRST current now reads\begin{eqnarray}
\bf {j}_{z}^{\textrm{B}} & = & -(\Pi _{zm}-\Pi ^{h}_{zm})\xi ^{m}-(id_{z\alpha }-id^{h}_{z\alpha })\lambda ^{\alpha }-(\partial \theta ^{\alpha }-\partial \theta ^{h\, \alpha })\chi _{\alpha }+\nonumber \\
& & +i\beta _{zm}(\lambda \gamma ^{m}\lambda )+2(\kappa _{z}\gamma _{m}\lambda )\xi ^{m} -\antifeld _{m}\xi '^{m}-\antifeld _{\q{\alpha }}\chi _{\alpha }'+b'_{zz}\gamma '^{z}+b_{zz}\gamma ^{z}+\nonumber \\
 &  & +c^{z}\Big (T_{zz}+\frac{1}{2}T_{zz}^{\textrm{top}}\Big )+\gamma ^{z}\Big (B_{zz}+\frac{1}{2}B_{zz}^{\textrm{top}}\Big)
\end{eqnarray}
with \begin{eqnarray}
T_{zz} & = & -\frac{1}{2}\Pi _{zm}\Pi _{z}^{m}+d_{z\alpha }\partial \theta ^{\alpha }+\frac{1}{2}\Pi ^{h}_{zm}\Pi _{z}^{h\, m}-d^{h}_{z\alpha }\partial \theta ^{h\, \alpha }+\beta _{zm}\partial \xi ^{m}+\omega _{z\alpha }\partial \lambda ^{\alpha }+\kappa ^{\alpha }_{z}\partial \chi _{\alpha }+\nonumber \\
 &  & +\beta '_{zm}\partial \xi '^{m}+\kappa '^{\alpha }_{z}\partial \chi '_{\alpha }+\antifeld _{z\mf{a}}\partial \feld ^{\mf{a}}+2\beta '_{zz}\partial \gamma '^{z}+\partial \beta '_{zz}\gamma '^{z}+2b'_{zz}\partial c'^{z}+\partial b'_{zz}c'^{z}\\
B_{zz} & = & -\frac{1}{2}(\Pi _{zm}+\Pi ^{h}_{zm})\beta_{z}^{m}+\frac{i}{2}(id_{z\alpha }+id^{h}_{z\alpha })\kappa ^{\alpha }_{z}-\frac{i}{2}(\partial \theta ^{\alpha }+\partial \theta ^{h\, \alpha })\omega _{z\alpha }+\nonumber \\
& & +\beta '_{zm}\partial \feld ^{m}+\kappa '^{\alpha }_{z}\partial \feld ^{\q{\alpha }}-2\beta '_{zz}\partial c'^{z}-c'^{z}\partial \beta '_{zz}-b'_{zz}-\frac{1}{2}c'^{z}F_{zzz}-\frac{1}{2}\gamma '^{z}\Phi _{zzz}
\end{eqnarray}
The final energy-momentum tensor, composite B-field and ghost current
are\begin{eqnarray}
{\bf T}_{zz} & = & T_{zz}+T^{\textrm{top}}=T_{zz}+2\beta _{zz}\partial \gamma ^{z}+\partial \beta _{zz}\gamma ^{z}+2b_{zz}\partial c^{z}+\partial b_{zz}c^{z}\\
{\bf B}_{zz} & = & B_{zz}+B^{\textrm{top}}=B_{zz}-2\partial c^{z}\, \beta _{zz}-c^{z}\partial \beta _{zz}\\
{\bf j}_{z}^{\textrm{gh}} & = & \beta _{zm}\xi ^{m}+\omega _{z\alpha }\lambda ^{\alpha }+\kappa _{z}^{\alpha }\chi _{\alpha }+\beta '_{zm}\xi '^{m}+\kappa _{z}'^{\alpha }\chi '_{\alpha }+\nonumber\\
& & +b'_{zz}c'^{z}+2\beta '_{zz}\gamma '^{z}+b_{zz}c^{z}+2\beta _{zz}\gamma ^{z}
\end{eqnarray}

Up to this point the above considerations were independent of the second
BRST operator. The straightforward definition $\mathbf Q_c=[\mathbf Q,K]$ 
with the simple homotopy $K$ that we used above, unfortunately, does not
yield a nilpotent $\mathbf Q_c$ because of quadratic and cubic antighost terms in the
generators $F_{zzz}$ and $\Phi_{zzz}$ in eqs. (\ref{Fzzz}) and (\ref{Phi}), whose contribution to $\mathbf Q$
does not vanish in $\mathbf Q_c^2=-\frac12[\mathbf Q,[K,[K,\mathbf Q]]]$.

\section{Conclusions and Outlook}

Starting from the classical Green Schwarz action we have constructed the
type II version of the covariant superstring of Grassi, Policastro, 
Porrati and van Nieuwenhuizen. In the first part of the paper we were aiming
at a transparent discussion of off-shell symmetries and their relation to
the on-shell constraint algebras. The gauging of the WZNW model was performed 
in an unconventional way that was guided by the cancellation of the central
terms in the constraint algebra. Circumventing the complications of the 
super group manifold approach we thus arrived at a free action with
a nilpotent BRST charge in a simple and straightforward way.

The next step was the construction of the second BRST charge $Q_c$ along the
lines suggested by Grassi and van Nieuwenhuizen in \cite{Nh:2004cz}. 
Such a charge is easily constructed using a homotopy, which implies
nilpotency due to the Jacobi identities.
A modification of our charge may, however, be necessary, because the 
difference to Berkovits' BRST operator is not yet $Q_c$ exact. 

One problem of the Berkovits approach that could be solved in 
\cite{Nh:2003kq} by adding the topological quartet is the absence of
diffeomorphism ghosts. With this quartet they also converted the Kazama
algebra into a twisted $N=2$ algebra. We observe, however, that one first
has to obtain $N=2$ with an additional quartet before 
diffeomorphisms (together with the accompanying 
fermionic symmetry) can be gauged. 

Once the free action and the BRST charges that define the relative
cohomology are constructed, it is clear that a number of tests
concerning the viability of this proposal should be made. 
Another crucial question is the possible
coupling of background fields, which again can be studied by cohomological
techniques of deformation theory. 
It would be quite 
interesting to gauge the world sheet translations and their fermionic
partner symmetry and thus obtain the topological quartet through a 
gauge fixing procedure of a diffeomorphism invariant theory on the world 
sheet.

\section*{Acknowledgements}

We would like to thank Peter van Nieuwenhuizen, who explained some
details of his and his collaborators' approach during his stay in
Vienna and Emanuel Scheidegger and Ulrich Theis for useful discussions.
S.G. would further like to thank Andreas Ipp for his frequent help
with the computer algebra program Mathematica. This work was partly
financed by the {}``Fonds zur F\"orderung der wissenschaftlichen
Forschung'' (FWF, project number P15553).

\newpage	\appendix

\section{Conventions}

 The main differences between our conventions and those of \cite{Nh:2003kq}
are:

\begin{eqnarray}
p & \To  & -p\\
d & \To  & -d\\
(\beta _{zm},\omega _{z\alpha },\kappa _{z}^{\alpha },b_{zz},\beta_{zz}) & \To  & (-\beta _{zm},-\omega _{z\alpha },-\kappa _{z}^{\alpha },-b_{zz},-\beta_{zz})\\
J_{M} & \neq  & (J_{m},J_{\alpha },J^{\alpha }),\quad J^{M}\neq (J^{m},iJ^{\alpha },-iJ^{\alpha })\\
\textrm{but }J_{M} & = & (J_{m},J_{\alpha },J_{\q{\alpha }})=(J_{m},J_{\alpha },-iJ^{\alpha })\\
J^{M} & = & (J^{m},J^{\alpha },J^{\q{\alpha }})=(J^{m},J^{\alpha },-iJ_{\alpha })\\
\mc {H}^{MN} & \To  & (-)^{N}\mc {H}^{MN}\\
f^{K}\tief{MN} & \To  & -f^{K}\tief{MN}
\end{eqnarray}
All other conventions are practically the same to make comparison
most simple. This includes a metric of the signature $(-,+)$ and the definition
of the lightcone coordinates in the way \begin{eqnarray}
\sigma ^{-} & \equiv  & \frac{1}{2}(\sigma ^{1}-\sigma ^{0})\quad \stackrel{\sigma ^{0}=-i\sigma ^{2}}{\longrightarrow }\quad \frac{1}{2}(\sigma ^{1}+i\sigma ^{2})\equiv z\\
\sigma ^{+} & \equiv  & \frac{1}{2}(\sigma ^{1}+\sigma ^{0})\quad \stackrel{\sigma ^{0}=-i\sigma ^{2}}{\longrightarrow }\quad \frac{1}{2}(\sigma ^{1}-i\sigma ^{2})\equiv \bar{z}\\
\partial \equiv \partial _{-} & = & \partial _{1}-\partial _{0}\quad \; \; \quad \stackrel{\sigma ^{0}=-i\sigma ^{2}}{\longrightarrow }\quad \partial _{1}-i\partial _{2}=\partial _{z}\equiv \partial \\
\bar{\partial }\equiv \partial _{+} & = & \partial _{1}+\partial _{0}\quad \; \; \quad \stackrel{\sigma ^{0}=-i\sigma ^{2}}{\longrightarrow }\quad \partial _{1}+i\partial _{2}=\partial _{\bar{z}}\equiv \bar{\partial }\\
g_{-+} & = & 2=g_{z\bar{z}}\\
g^{-+} & = & \frac{1}{2}g^{z\bar{z}}
\end{eqnarray}
The right hand side is the definition of the complex coordinates of
the Euclidean theory. We do not distinguish between \( \partial _{-} \)
and \( \partial _{z} \) and call both of them \( \partial  \). The
conformal map of the closed string worldsheet to the punctured complex
plane has to look as follows\begin{eqnarray}
z' & = & e^{-2iz}=e^{-i\sigma ^{1}+\sigma ^{2}}
\end{eqnarray}
We will not distinguish between \( z \) and \( z' \). In OPE's,
the variable \( z \) is the one of the punctured plane.

Wherever the string parameter \( \alpha ' \) is suppressed, it is
set to \begin{equation}
\alpha '=2
\end{equation}
For the representation of the Gamma matrices, we refer to \cite[p.14]{Nh:2003cm}.
All what we need from the discussion in there is\begin{eqnarray}
C\Gamma ^{m} & = & \left( \begin{array}{cc}
\gamma ^{m}_{\alpha \beta } & 0\\
0 & \gamma ^{m\, \alpha \beta }
\end{array}\right) 
\end{eqnarray}
With \emph{symmetric} matrices \( \gamma ^{m}_{\alpha \beta } \)
and \( \gamma ^{m\, \alpha \beta } \) which both obey (up to the
position of the indices) a \textbf{Fierz identity} of the form\begin{equation}
\gamma _{m(\alpha \beta }\gamma ^{m}_{\gamma )\delta }=0
\end{equation}
In addition the following equation holds\begin{equation}
\gamma ^{m}_{\alpha \beta }\gamma ^{n\, \beta \gamma }+\gamma ^{n}_{\alpha \beta }\gamma ^{m\, \beta \gamma }=-2\eta ^{mn}\delta ^{\gamma }_{\alpha }
\end{equation}
(Here we disagree with the sign of \cite[p.16]{Nh:2003cm}). For a
Dirac spinor \( \Psi =(\psi ^{\alpha },\varphi_{\alpha }) \),
we thus have \begin{eqnarray}
\bar{\Psi }\Gamma ^{m}\Psi  & = & \psi ^{\alpha }\gamma ^{m}_{\alpha \beta }\psi ^{\beta }+\varphi_{\alpha }\gamma ^{m\, \alpha \beta }\varphi_{\beta }\equiv (\psi \gamma ^{m}\psi )+(\varphi\gamma ^{m}\varphi)
\end{eqnarray}
We will consider Majorana-Weyl Femions only. \( \theta ^{\alpha } \)
e.g. is real and has \( 16 \) components. A hat \( \widehat{\: } \)
on the index allows to treat type IIA and type IIB strings at the
same time:\begin{equation}
\hat{\theta }^{\hat{\alpha }}\equiv \left\{ \begin{array}{c}
\hat{\theta }_{\alpha }\quad \textrm{for type IIA}\\
\hat{\theta }^{\alpha }\quad \textrm{for type IIB}
\end{array}\right. 
\end{equation}
In the text we often call the hatted variables ``antichiral'' and the others
``chiral''. This refers
to ``right-moving'' and ``left-moving'' and does not necessarily imply type
IIA.

\subsection{Minkowskian and Euclidean Lagrangian}

The \textbf{Minkowskian} free field action is \begin{eqnarray}
S & = & \frac{1}{2\pi \alpha '}\int \de ^{2}\sigma \sqrt{-g}\underbrace{\left( -\frac{1}{2}\partial _{\mu }x^{m}\partial ^{\mu }x_{m}+P^{\mu \nu }p_{\mu \alpha }\partial _{\nu }\theta ^{\alpha }\right) }_{\equiv \mc {L}}\\
\de ^{2}\sigma  & = & \de \sigma ^{0}\de \sigma ^{1}\\
P^{\mu \nu } & \equiv  & g^{\mu \nu }-\eps ^{\mu \nu },\quad \eta _{\mu \nu }=\diag (-1,1)\\
\eps ^{\mu \nu } & \equiv  & \frac{\epsilon ^{\mu \nu }}{\sqrt{-g}},\quad \epsilon ^{01}=1=-\epsilon ^{10}
\end{eqnarray}
In our calculations we consider coordinates \( \sigma ^{0} \) and
\( \sigma ^{1} \) for which we have a flat worldsheet metric \( \eta ^{\mu \nu } \)
and stick to this \( \sigma ^{0} \) as the canonical time. This implies
that we do not transform the measure, when we consider the Lightcone
coordinates. We denote by \( \mc {L} \) not the complete Lagrangian
but only the inner (scalar) part\begin{equation}
\mc {L}\equiv -\frac{1}{2}\partial _{\mu }x^{m}\partial ^{\mu }x_{m}+P^{\mu \nu }p_{\mu \alpha }\partial _{\nu }\theta ^{\alpha }
\end{equation}
So if we change to lightcone coordinates, we do not have to consider
a factor of \( 2 \) coming from \( \sqrt{-g} \) . One can then just
write an equality of the form\begin{eqnarray}
\mc {L} & \equiv  & -\frac{1}{2}\partial _{\mu }x^{m}\partial ^{\mu }x_{m}+P^{\mu \nu }p_{\mu \alpha }\partial _{\nu }\theta ^{\alpha }=\\
 & = & -\frac{1}{2}\partial x^{m}\bar{\partial }x_{m}+p_{-\alpha }\bar{\partial }\theta ^{\alpha }
\end{eqnarray}
We arrive at the \textbf{Euclidean} action by replacing the Minkowskian
by an Euclidean metric, multiplying with an overall minus sign and
redefining \( \eps ^{\mu \nu } \) with an extra \( -i \).\begin{eqnarray}
S^{E} & = & -\frac{1}{2\pi \alpha '}\int \de ^{2}\sigma \sqrt{g}\underbrace{\left( -\frac{1}{2}\partial _{\mu }x^{m}\partial ^{\mu }x_{m}+P^{\mu \nu }p_{\mu \alpha }\partial _{\nu }\theta ^{\alpha }\right) }_{\equiv \mc {L}^{E}}\\
\de ^{2}\sigma  & = & \de \sigma ^{2}\de \sigma ^{1}\\
P^{\mu \nu } & \equiv  & g^{\mu \nu }-\eps ^{\mu \nu }\\
\eps ^{\mu \nu } & \equiv  & -\frac{i\epsilon ^{\mu \nu }}{\sqrt{g}},\quad \epsilon ^{12}=1=-\epsilon ^{21}
\end{eqnarray}
For a flat metric we thus do not have to distinguish between Minkowskian
and Euclidean Lagrangian, so we will use the indices {}``\( - \)''
and {}``\( z \)'' synonymously.

Switching to \textbf{complex coordinates} in some sense undoes the
Wick rotation as the determinant of the metric becomes negative. The
measure transforms as follows\begin{eqnarray}
d^{2}z & \equiv  & \de z\de \bar{z}=\frac{1}{4}\left( \de \sigma ^{1}+i\de \sigma ^{2}\right) \left( \de \sigma ^{1}-i\de \sigma ^{2}\right) =\frac{i}{2}\de \sigma ^{2}\de \sigma ^{1}=\frac{i}{2}\de ^{2}\sigma \\
\de ^{2}\sigma  & = & -2i\de ^{2}z
\end{eqnarray}
\begin{equation}
S_{\textrm{comp}}^{E}=\frac{i}{\pi \alpha '}\int \de ^{2}z\underbrace{\left( -\frac{1}{2}\partial x^{m}\bar{\partial }x_{m}+p_{z\alpha }\bar{\partial }\theta ^{\alpha }\right) }_{\mc {L}_{\textrm{comp}}^{E}}
\end{equation}
\begin{eqnarray}
\eps ^{z\bar{z}} & = & -\frac{1}{2}=\eps ^{-+}\\
\eps _{z\bar{z}} & = & 2=\eps _{-+}
\end{eqnarray}
In explicit calculations we do not specify whether we are in the Euclidean or in
the Minkowskian case
and thus just write\begin{equation}
S=\int \mc {L}
\end{equation}
where all the necessary prefactors and the measure are part of the
\( \int  \)-sign. As it looks nicer, we will write the index \( z \)
whenever we are either in lightcone or in complex coordinates. We treat
these coordinates in the same way, as already mentioned.

\subsection{Superspace-Conventions}

Similar to \cite{Nh:2003kq}, we use strict Southwest-Northeast conventions
(NE for short) for capital indices, i.e. they are contracted in the
following way:\begin{equation}
T_{M}J^{M}=(-)^{M}J^{M}T_{M}
\end{equation}
In this convention it also makes sense to introduce two different
Kronecker-deltas:\begin{eqnarray}
\delta ^{M}_{N}\equiv \delta ^{M}\tief{N} & = & (-)^{MN}\delta _{N}\hoch{M}=(-)^{M}\delta _{N}\hoch{M}
\end{eqnarray}
where the lefthand side is numerically equal to the usual Kronecker
delta. However, we do not make this distinction for small indices,
because they are always of definite grading\begin{equation}
\delta _{\alpha }^{\beta }=\delta _{\alpha }\hoch{\beta }=\delta ^{\beta }\tief{\alpha }
\end{equation}
Also the contraction direction (NE) is not important for the small
indices. For matrices we define a \textbf{graded inverse} that yields
the appropriate Kronecker delta. In the case of a metric the graded
inverse gets the same symbol with different index positions:\begin{eqnarray}
\mc {H}_{MP}\mc {H}^{PN} & = & \delta _{M}\hoch{N}=(-)^{M}\delta _{M}^{N}\\
(-)^{P}\mc {H}^{MP}\mc {H}_{PN} & = & \delta ^{M}\tief{N}=\delta ^{N}_{M}
\end{eqnarray}
From the second line one sees how the graded inverse is numerically
related to the ordinary inverse:\begin{eqnarray}
(\mc {H}^{-1})^{MP} & = & (-)^{P}\mc {H}^{MP}
\end{eqnarray}
We denote the graded commutator always with the ordinary squared
brackets:
\begin{equation}[A,B]\equiv AB-(-)^{AB}BA\end{equation}

\rem{
\section{OPEs}
\section{WZNW}
\begin{itemize}
\item not free fields
\item two sectors already for heterotic string
\item 4 sectors for type II string, Jacobi-identities
\item not off-shell supersymmetric for type II
\end{itemize}
}

\vspace{0.5cm}
\begingroup
\raggedright

\endgroup


\end{document}